\documentclass[twocolumn]{aastex62}

\usepackage{graphicx}
\usepackage{amssymb}
\usepackage{amsmath}
\usepackage{natbib}
\usepackage{wrapfig}
\usepackage[utf8]{inputenc}
\usepackage[T1]{fontenc}
\shorttitle{SMC Kinematics}
\shortauthors{Zivick et al.}

\begin{document}

\title{The Proper Motion Field of the Small Magellanic Cloud: Kinematic Evidence for its Tidal Disruption}

\correspondingauthor{Paul Zivick}
\email{pjz2cf@virginia.edu}

\author[0000-0001-9409-3911]{Paul Zivick}
\affiliation{Department of Astronomy, University of Virginia, 530 McCormick Road, Charlottesville, VA 22904, USA}

\author{Nitya Kallivayalil}
\affiliation{Department of Astronomy, University of Virginia, 530 McCormick Road, Charlottesville, VA 22904, USA}

\author{Roeland P. van der Marel}
\affiliation{Space Telescope Science Institute, 3700 San Martin Drive, Baltimore, MD 21218, USA}
\affiliation{Center for Astrophysical Sciences, Department of Physics \& Astronomy, Johns Hopkins University, Baltimore, MD 21218, USA}

\author{Gurtina Besla}
\affiliation{Steward Observatory, University of Arizona, 933 North Cherry Avenue, Tucson, AZ 85721, USA}

\author{Sean T. Linden}
\affiliation{Department of Astronomy, University of Virginia, 530 McCormick Road, Charlottesville, VA 22904, USA}

\author{Szymon Koz{\l}owski}
\affiliation{Warsaw University Observatory Al. Ujazdowskie 400-478 Warszawa, Poland}

\author{Tobias K. Fritz}
\affiliation{Department of Astronomy, University of Virginia, 530 McCormick Road, Charlottesville, VA 22904, USA}
\affiliation{Instituto de Astrofisica de Canarias, Calle Via Lactea s/n, E-38205 La Laguna, Tenerife, Spain}
\affiliation{Universidad de La Laguna, Dpto. Astrofisica, E-38206 La Laguna, Tenerife, Spain}

\author{C. S. Kochanek}
\affiliation{Department of Astronomy, The Ohio State University, 140 West 18th Avenue, Columbus, OH 43210, USA}
\affiliation{Center for Cosmology and AstroParticle Physics, The Ohio State
University, 191 W. Woodruff Ave., Columbus, OH 43210, USA}

\author{J. Anderson}
\affiliation{Space Telescope Science Institute, 3700 San Martin Drive, Baltimore, MD 21218, USA}

\author{Sangmo Tony Sohn}
\affiliation{Space Telescope Science Institute, 3700 San Martin Drive, Baltimore, MD 21218, USA}

\author{Marla C. Geha}
\affiliation{Department of Astronomy, Yale University, New Haven, CT 06520, USA}

\author{Charles R. Alcock}
\affiliation{Harvard-Smithsonian Center for Astrophysics, 60 Garden Street, Cambridge, MA 02138, USA}

\begin{abstract}

We present a new measurement of the systemic proper motion of the Small Magellanic Cloud (SMC), based on an expanded set of 30 fields containing background quasars and spanning a $\sim$3 year baseline, using the \textit{Hubble Space Telescope} (\textit{HST}) Wide Field Camera 3. Combining this data with our previous 5 \textit{HST} fields, and an additional 8 measurements from the \textit{Gaia}-Tycho Astrometric Solution Catalog, brings us to a total of 43 SMC fields. We measure a systemic motion of $\mu_{W}$ = $-0.82$ $\pm$ 0.02 (random) $\pm$ 0.10 (systematic) mas yr$^{-1}$ and $\mu_{N}$ = $-1.21$ $\pm$ 0.01 (random) $\pm$ 0.03 (systematic) mas yr$^{-1}$. After subtraction of the systemic motion, we find little evidence for rotation, but find an ordered mean motion radially away from the SMC in the outer regions of the galaxy, indicating that the SMC is in the process of tidal disruption. We model the past interactions of the Clouds with each other based on the measured present-day relative velocity between them of $103 \pm 26$ km s$^{-1}$. We find that in 97\% of our considered cases, the Clouds experienced a direct collision $147 \pm 33$ Myr ago, with a mean impact parameter of $7.5 \pm 2.5$ kpc.
\end{abstract}

\section{Introduction} \label{sec:introduction}

Our understanding of the Small and Large Magellanic Clouds (SMC and LMC) has evolved greatly in the age of space-based proper motion (PM) measurements. The \textit{HST} PM measurements by \cite{NK06a} were used to demonstrate that the Clouds had not orbited the Milky Way (MW) multiple times as expected but instead were likely on their first infall into the MW \citep{besla07}. With the supporting results from \cite{piatek08}, this view of the Clouds became the new paradigm and has driven our understanding of their evolution.

Since then, the evolution of the LMC has proved more tractable to understanding. Using a longer baseline and the then new Wide Field Camera 3 (WFC3), Kallivayalil et al. (2013, hereafter NK13) significantly improved the PM errors for 26 LMC fields. Using the decreased uncertainties, \cite{vdM14} were able to make a direct measurement of the PM rotation curve of the LMC in the plane of the sky. A followup examination of the center of mass (COM) PM of the LMC and its rotation curve using PMs from the Tycho-\textit{Gaia} Astrometric Solution (TGAS) Catalog \citep{lindegren16}, which combines \textit{Gaia} Data Release 1 \citep{gaiadr1a,gaiadr1b} with the \textit{Hipparcos} Tycho-2 Catalog \citep{hoeg00}, supported this finding (\citealt{vdM16}; hereafter vdMS16), suggesting that the inner region of the LMC is a relatively well-behaved system. Further out the picture becomes more complicated with increasing evidence for more complicated substructures in the periphery of the LMC (\citealt{choi18a, choi18b}; \citealt{mackey18}; \citealt{nidever18}).

The structure and dynamics of the SMC has not proved to be as easy to understand. NK13 had results for only five fields, enough to attempt a measurement of the COM PM, but not enough to describe the internal kinematics. vdMS16 analyzed PMs for eight individual stars in the SMC from the TGAS Catalog \citep{lindegren16}, but the resulting residual motions were not indicative of any coherent motion. A third COM PM was measured by \cite{cioni16} as a by-product of their work on 47 Tuc did not provide any additional insight into the internal workings of the SMC.

Line-of-sight (LOS) motion studies have attempted to fill this gap. \cite{stanimirovic} found a rotation signature in the H {\small I} gas in the SMC with the line of nodes, defined as the line joining the points of maximum and minimum relative velocity, parallel to the visible major axis of the SMC and a dynamical center located in the northeastern section of the SMC. A study of the red giants in the SMC by \cite{harris06} suggested that the older population was dynamically separate from the neutral gas, having a very weak rotation signature and a much more significant velocity dispersion, suggesting a spheroidal rather than disk structure for the SMC. However, \cite{dobbie14a} conducted a broader investigation of the red giant population, extending beyond the central area considered by \cite{harris06}, and found instead a rotation signature of 20--40 km s$^{-1}$ kpc$^{-1}$, although their line of nodes did not agree with that found by \cite{stanimirovic}. To complicate the picture further, \cite{evans08} found a similar rotation curve to that of \cite{dobbie14a} but in the young, massive star population (O, B, and A stars). A slight velocity gradient was also found for the OB stars by 
\cite{lamb16}. While one could argue that the red giant population could be dynamically decoupled from the underlying neutral gas, one would not expect the same to have happened for the young stars. \cite{dobbie14a} proposed an inclined disk to help explain the differences, but further kinematic evidence is needed to fully evaluate this possibility.

Because of the interest in the nature of the past mutual interactions of the Clouds, many photometric studies have searched for tidal debris at large radii from the Clouds, or evidence for SMC stars in the stream of H {\small I} gas linking the LMC and SMC, referred to as the Magellanic Bridge. There is evidence for both young stars, presumably formed \textit{in situ} \citep{harris07}, and intermediate-age stars in the Bridge (\citealt{bagheri13}; \citealt{nidever13}; \citealt{skowron14}; \citealt{noel15}). Even old stellar populations have been detected in the vicinity of, but not aligned with, the gaseous Bridge (\citealt{belokurov16}; \citealt{deason16}; \citealt{belokurov17}; \citealt{carrera17}), adding support to the idea that the SMC is being tidally stripped by the LMC.




\cite{dias16} studied the ages of star clusters throughout the SMC, finding clear age and metallicity gradients consistent with
tidal interactions between the LMC and SMC. These results are supported by numerical models of the Magellanic system that predict that the SMC should be constantly churning and only at large radii would there potentially be a coherent rotation signature \citep{besla12}. A comprehensive study of the classical Cepheids in the SMC \citep{ripepi17} found a complex geometric structure, with the near side forming a rough spheroidal shape before gradually shifting into a more linear shape, adding more detail to the SMC but presenting yet another potentially conflicting stellar structure.

In order to address the still sparse PM coverage of the SMC, we obtained two epochs of \textit{HST} data with WFC3/UVIS for 30 fields in the SMC, focusing on obtaining data in the outer regions. Combined with the data already available from NK13 and vdMS16 for the interior of the SMC, this provides a broader kinematic view of the SMC. 
In this paper, we present the results from this combined dataset, and their implications for the orbital history of the SMC. 

The paper is organized as follows. In Section \ref{sec:data} we discuss the quality of the data, our process for creating an astrometric reference frame, and how we quantify the uncertainties in the PMs for each field. We use these motions in Section \ref{sec:pmresults}, in conjunction with a dynamical model similar to the one in \cite{vdM02}, to analyze the measured PMs under various model assumptions. This produces a set of best fit parameters, including measurements of the overall COM PM for the SMC and a measurement of the rotational velocity. In Section \ref{sec:internal}, we 
subtract the best fit COM motion to study the
 internal motions of the SMC, both for all stars and by stellar type. 
We use the newly determined COM PM to constrain the SMC's past orbit, and examine its interactions with the LMC and MW in Section \ref{sec:orbit}. Finally, in Section \ref{sec:disc} we discuss the overall ramifications of the new SMC data for our understanding of the Magellanic system and where future studies will allow for further improvements.

\section{Data and Analysis} \label{sec:data}
\subsection{Description of Observations} \label{sec:obs}
\begin{figure}
\begin{center}
 \includegraphics[width=3.4in]{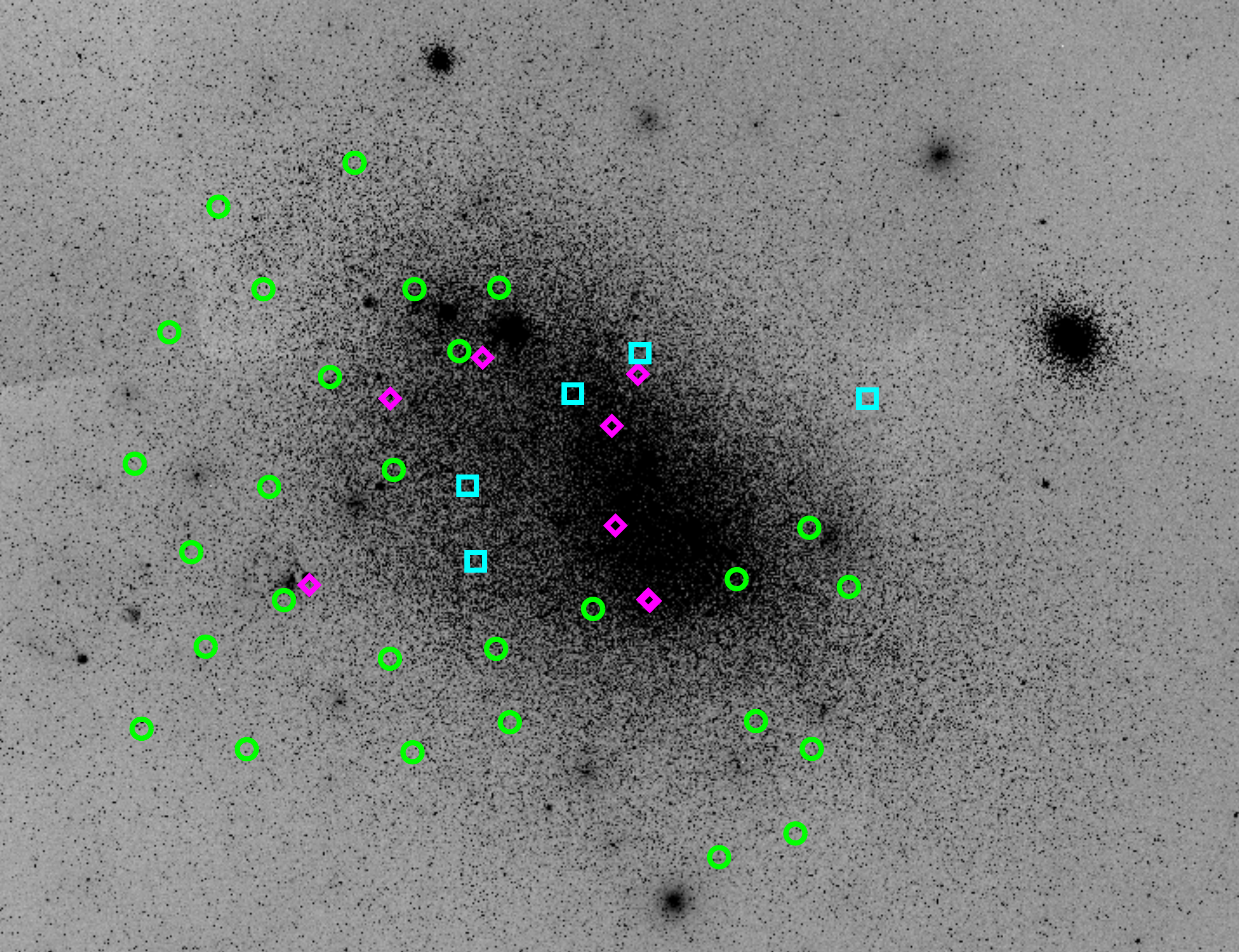} 
 \caption{B-band image covering 4$^{\circ} \times 6^{\circ}$ from the Digital Sky Survey of the SMC (with 47 Tuc to the right) where north is up and east is to the left in the image. The LMC, northeast of the SMC, is located off the panel to the upper left. The green circles indicate the locations of our new reference QSOs from \cite{kozlowski13}. The cyan squares show the quasars used in NK13 and \cite{NK06a}, and the magenta diamonds indicate the positions of the stars used in vdMS16.}
   \label{fig:smcpos}
\end{center}
\end{figure}

\cite{kozlowski13} spectroscopically confirmed the presence of nearly 200 QSOs behind the SMC. From this set of QSOs, we selected 30 of the brightest QSOs ($17.7 \leq V \leq 20.1$ mag) to provide an inertial reference frame across the two epochs of observations. The QSOs were also selected to provide roughly uniform coverage of the SMC, over an approximately 16 square degree patch of sky (see Figure \ref{fig:smcpos}). Such a uniform sampling was required to better sample the kinematic behavior of the SMC. \cite{kozlowski13} were unable to observe their candidates in the NW corner of the SMC, leaving us with no spectroscopically confirmed QSOs to target, limiting our target fields to the central body and the southern and eastern periphery.

Both epochs of data were collected with the \textit{HST} WFC3/UVIS, with the first epoch beginning observations in 2013 and the second epoch beginning in 2016, to provide a roughly 3 year baseline for each field (see Table \ref{tab:pm}). In the first epoch, four observations were collected with the F606W filter using a custom DITHER-BOX pattern to provide for optimal sampling of the point-spread function (PSF). The exposure times for each field ranged in length from 2-6 minutes to achieve a signal-to-noise of $\geq$200 for the QSOs. Two additional short-exposure observations were obtained with the F814W filter to make color magnitude diagrams (CMDs), to separate SMC and field stars, and to assist in identifying the QSO (see Section \ref{sec:obschar}). For the second epoch, six dithered observations were collected with the F606W filter, and no observations were made with F814W filter as the astrometric transformations were only to be made with the F606W data. The orientation of the instrument was required to be the same for both observations of each field in order to minimize systematic errors.

\subsection{Analysis of WFC3/UVIS Observations} \label{sec:obschar}

For our analysis, we used the bias-subtracted, dark-subtracted, flat-fielded, and CTE-corrected images (\_flc.fits) provided by the Space Telescope Science Institute (STScI) data reduction pipeline. The individual dithered images provide better astrometry than the standard MultiDrizzle data product (\_drc.fits), as noted in \cite{anderson04}. However, unlike the drizzled images, these data are not corrected for  geometric distortion. To address this, we apply the known geometric distortion solution for WFC3/UVIS \citep{bellini11} to the positions of the sources rather than correcting the images. These positions were measured using an empirically built PSF library for WFC3/UVIS, constructed similarly to the process described in \cite{anderson06}.

Once a list of sources was created from each individual image the pixel positions were converted into the WCS frame using the information contained within the \_flc headers. The 30 brightest objects were selected from each list and matched using the WCS solution from the headers with a healthy tolerance of 20 arcseconds. This tolerance was chosen after a series of manual trials. 
We then used the brightest six matched objects 
as the initial constraints to linearly transform all sources
into the reference frame of the first epoch.

\begin{figure}
\begin{center}
 \includegraphics[width=3.4in]{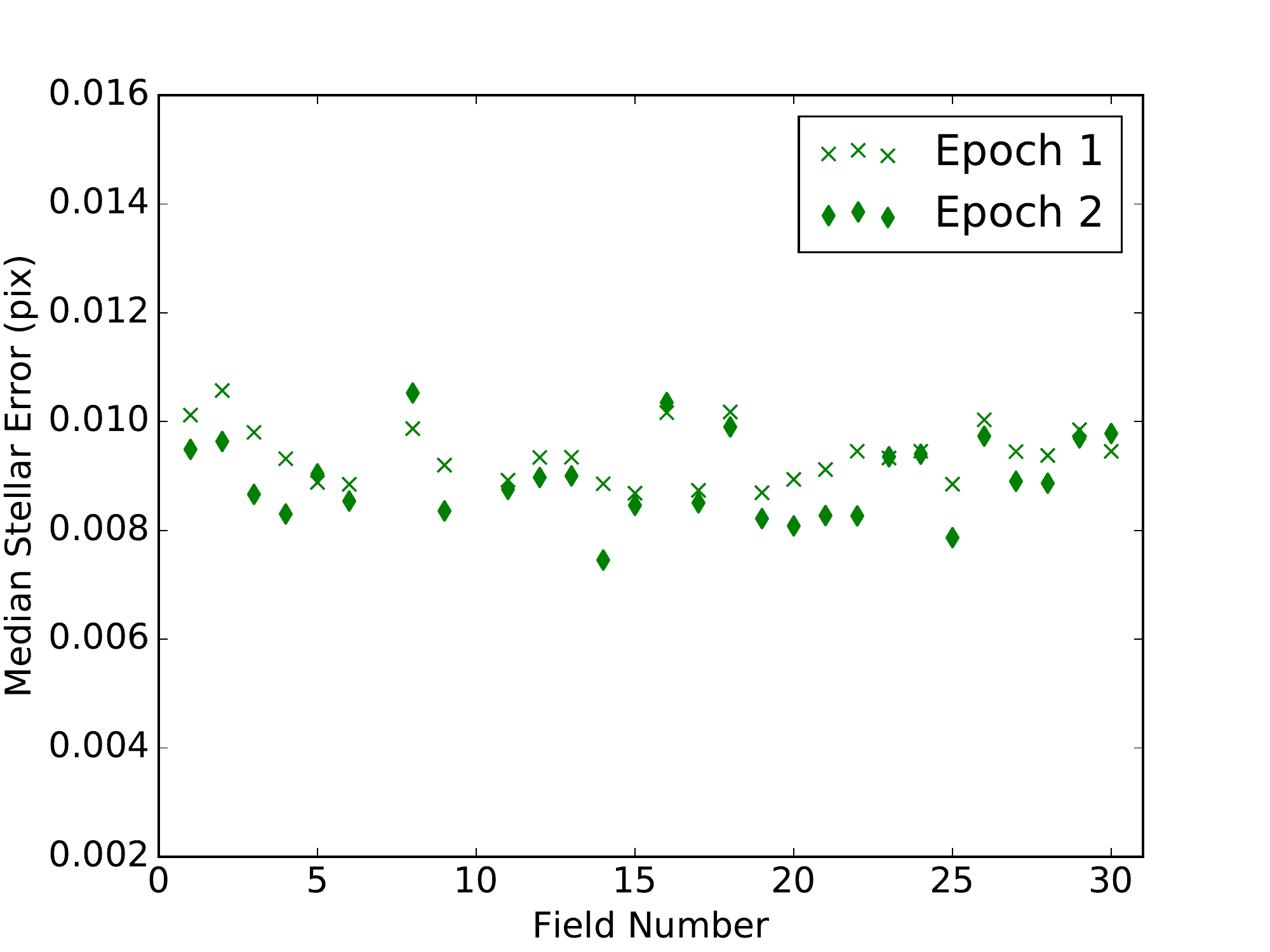} 
 \caption{Median standard deviation of the source positions for all fields used in the analysis. The green points represent the median of all stellar sources in a field. The crosses represent measurements for the first epoch of observations and the diamonds represent the second epoch.}
   \label{fig:mederr}
\end{center}
\end{figure}

\begin{figure}
\begin{center}
 \includegraphics[width=3.4in]{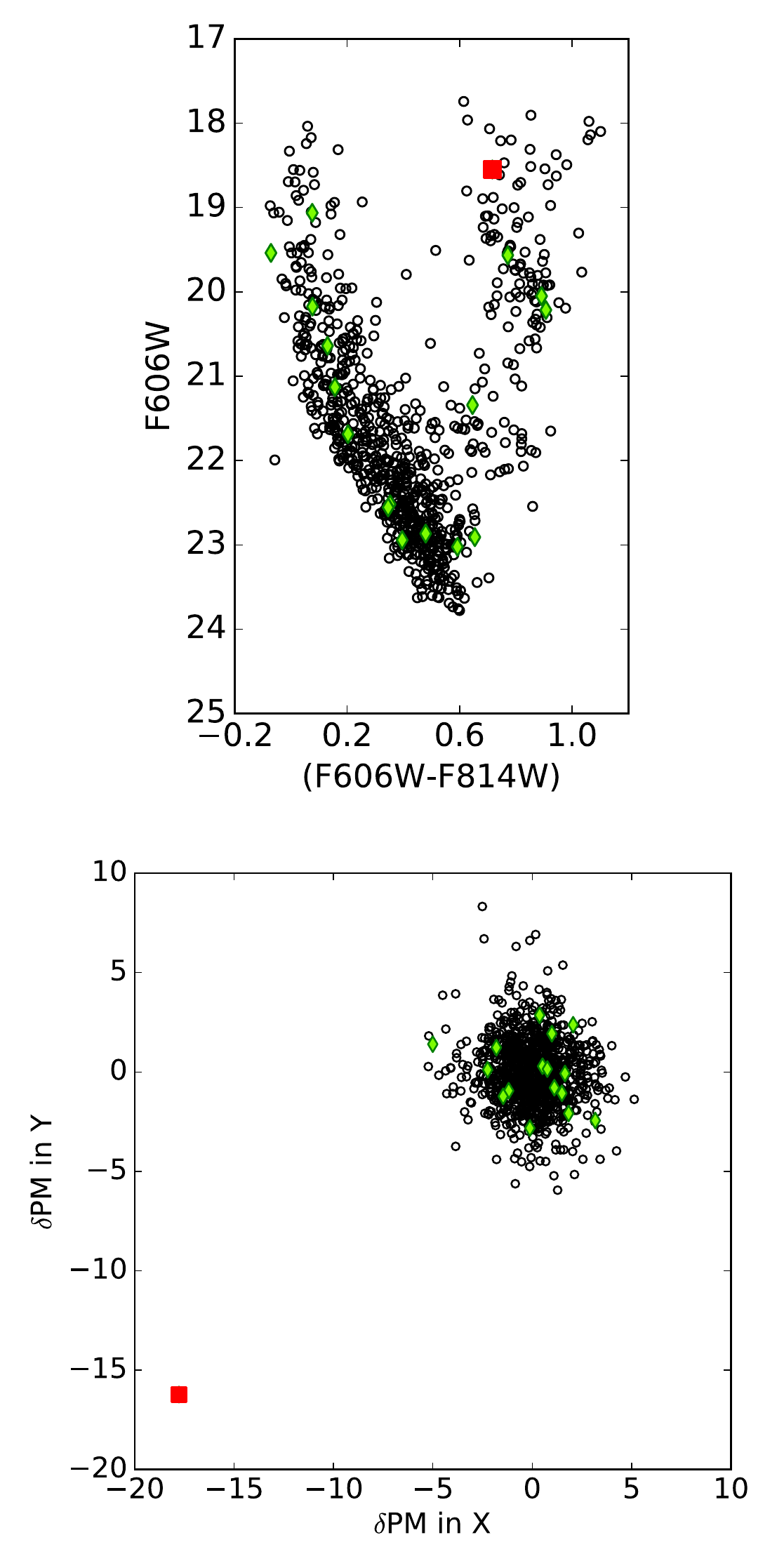} 
 \caption{Example of all sources in the vicinity of the known position of the quasar. Black points represent all sources in the field, green diamonds represent the sources within 7 arcseconds of the known QSO position, and the red square represents the QSO. (Top) CMD for the field. (Bottom) PM divided by the scatter in position in the pixel frame, described further in Section \ref{sec:twoepoch}.}
   \label{fig:f6qsoprop}
\end{center}
\end{figure}

Within a given field, the number of common sources varied from roughly 100 in the sparse fields toward the outer edges of the SMC to more than 1000 in the fields closer toward the visible body of the SMC. For the final transformations and iterations, we required that every source was detected in all 10 dithered images, 4 from the first epoch and 6 from the second epoch, to simplify the uncertainty estimate.

The positional errors of the matched stars increase slowly as a function of magnitude, beginning at F606W $\sim 20$. For the iterative linear transformations a minimum error of 0.005 pixels was added to all sources brighter than 20 mag, measured from the median scatter for all sources with $18 <$ F606W $< 20$, in order to avoid overweighting the transformation. The median scatter is roughly consistent across all fields (as seen in Figure \ref{fig:mederr}).


The next step is to identify the QSO. 
Using the known location of the QSO, we select all candidate objects within several arcseconds. From there, using a combination of the photometric and kinematic properties of the objects, we are able to identify the QSO, as demonstrated in Figure \ref{fig:f6qsoprop}. Note that in our analysis, the SMC stars have zero average motion by construction, so the reflex motion of the QSO with respect to the stars is our measured signal, as can be seen with the red point in Figure \ref{fig:f6qsoprop}. For Fields 7 and 10 we were unable to measure the motion of the QSOs.  For Field 7, a foreground star overlapped the QSO, and for Field 10, the host galaxy was resolved, both situations causing a large scatter in position. In the second case, the host galaxy of the QSO was resolved and the uncertainty from fitting the galaxy with a Sersic profile and point source was larger than the expected PM signal.


\begin{figure}
\begin{center}
 \includegraphics[width=3.4in]{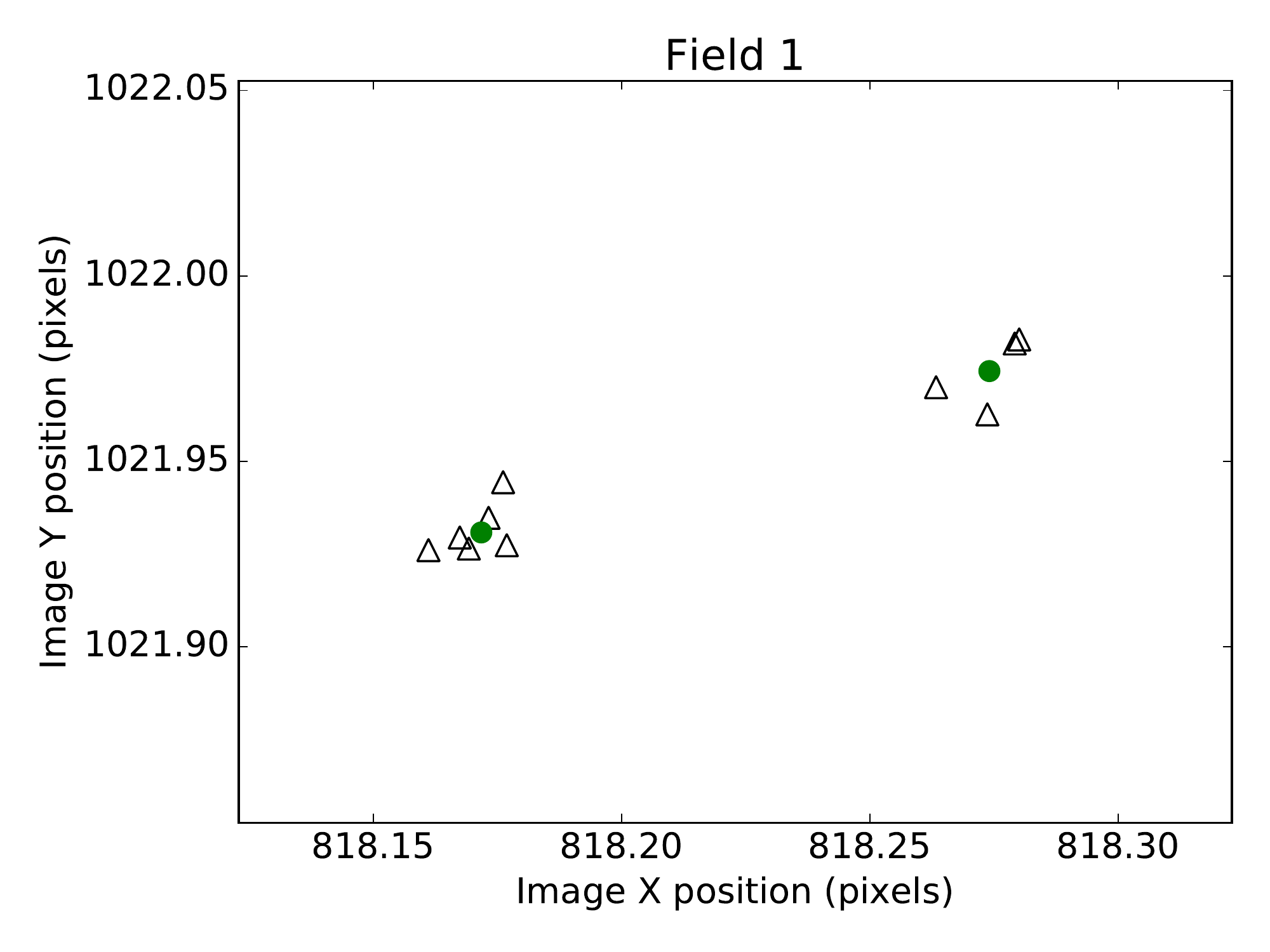} 
 \caption{Positions in the master reference frame of the QSO from each individual dither (black triangles) for Field 1, with the average for each epoch marked by green circle. The "motion" is the inverse of the average motion of the stars, as we measure the QSO position relative to the stellar frame.}
   \label{fig:dithers}
\end{center}
\end{figure}

\subsection{Two Epoch Results} \label{sec:twoepoch}

Once the initial transformation using the first six common sources was performed, each iteration thereafter applied stricter constraints on which sources were to be considered SMC sources. For each source, all dither positions in a given epoch were averaged together with the standard deviation of the positions used as the estimate for the positional uncertainty. The averaged positions between the epochs were subtracted from each other, with this difference then divided by the total error of the source (the standard deviation from each epoch added in quadrature) to create the $\delta$PM for the source, a measure of the statistical significance of the motion. Using these two measurements, thresholds were iteratively decreased, beginning with a one pixel tolerance and a $\delta$PM of 50 and ending with requiring all used sources to move less than 0.1 pixel and have a $\delta$PM less than 5, over an average of five steps. 

\begin{figure}
 \includegraphics[width=3.4in]{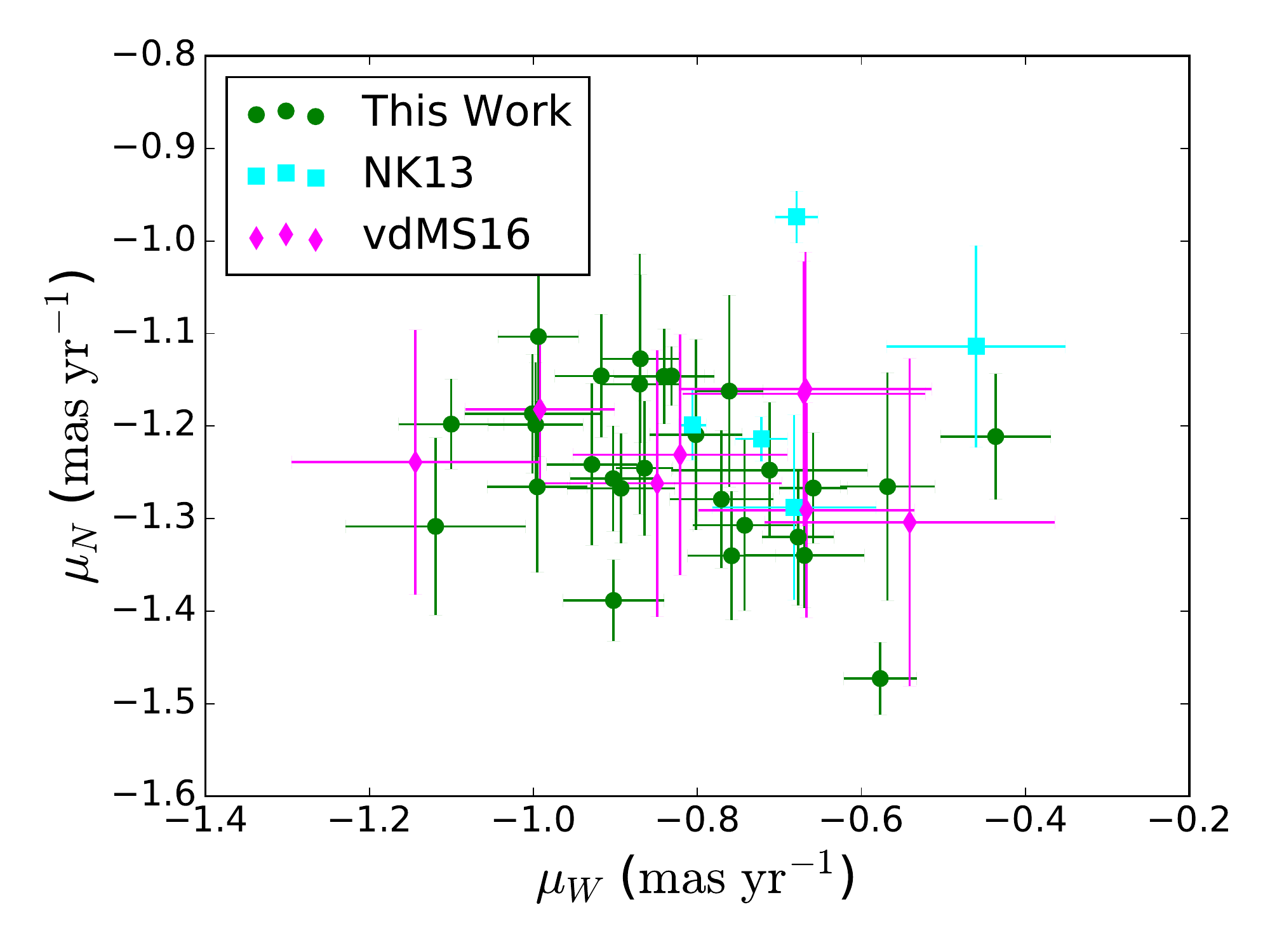} 
 \caption{The measured PMs for each field (green), along with the earlier measurements by NK13 (cyan) and vdMS16 (magenta).}
   \label{fig:errmot}
\end{figure}

The final motion for the field is the difference of the average positions of the QSO in the two epochs with the scatter from the individual images providing an estimate of the uncertainties. Figure \ref{fig:dithers} shows an example of the quasar in Field 1. This difference is then divided by the baseline for the observations of the field, converted to milliarcseconds (mas), and decomposed into the WCS frame using the position angle for the observation and then inverted to provide the motion of the stars, rather than the reflex motion of the QSO. We defined a local reference frame of $\mu_{W}$ and $\mu_{N}$ to account for the impact of declination ($\delta$) on the apparent motion in right ascension ($\alpha$), with $\mu_{W} \equiv -(d\alpha/dt) cos(\delta)$ and $\mu_{N} \equiv d\delta/dt$. The resulting PMs for all successfully measured fields, as well as prior PM measurements for the SMC, are shown in Figure \ref{fig:errmot}.

For the final error estimate in each field, we begin with the error in the pixel frame, which has two components. The first component is the standard deviation of the QSO positions, with the error for each epoch added in quadrature, $\delta$PM$_{\mathrm{QSO}}$. The second component is the scatter in the difference in position between epochs for all stars used in the transformation, $\sigma_{\langle PM \rangle}$. By construction, this value should be zero as the stars are aligned to themselves, so the deviation from zero acts as an estimate for the accuracy of the transformation. These two components were added together in quadrature and then converted to $\mu_{W}$ and $\mu_{N}$, including a covariance term for the errors to account for the rotation relative to the pixel frame.

\begin{figure}
\begin{center}
 \includegraphics[width=3.4in]{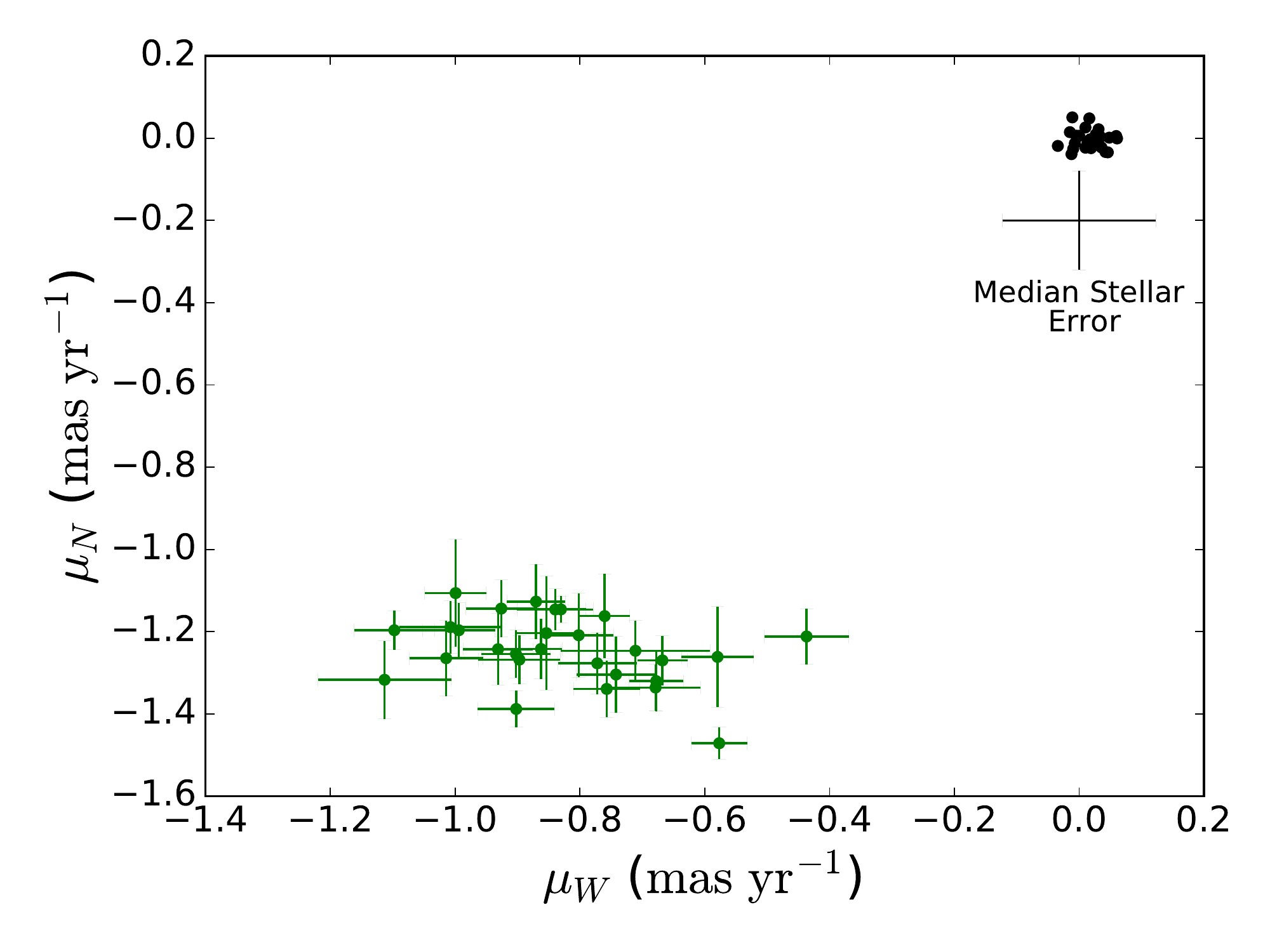} 
 \caption{The motions inferred for each quasar (green points), corrected for the reflex motion, as compared to the median motions for the stars (black points). The median uncertainty for the stars is shown. That the stars cluster close to zero and that the average of the stars is still closer to zero indicates that the transformations have worked as intended.}
   \label{fig:qsostellar}
\end{center}
\end{figure}

Figure \ref{fig:qsostellar} demonstrates that the transformations worked as intended. The stellar motions cluster around zero (with the median stellar error displayed below the cluster for reference), and the motions derived from the QSOs clearly separate from the motions of the SMC member stars in each field relative to one another.

\begin{table*}[t]
\begin{center}
\caption{SMC Observations and Results.\label{tbl-2}}
\begin{tabular}{lllllllcrcccc}
\tableline
\tableline
 & & & & & & & \multicolumn{3}{c}{New SMC PMs} & & &  \\ \cline{6-10}

ID & \multicolumn{3}{c}{R.A.} & \multicolumn{3}{c}{Decl.} & $\Delta$ Time & \textit{N} & \multicolumn{4}{c}{PM of Field as Observed} \\ \cline{10-13}

& & & & & & & & & $\mu_{W}$ & $\mu_{N}$& $\delta\mu_{W}$& $\delta\mu_{N}$ \\ \cline{10-13}

 & h & m & s & deg & ' & '' & (years) & & \multicolumn{4}{c}{(mas yr$^{-1}$)} \\

\tableline
SZ1 & 0 & 37 & 4.7 & $-$73 & 22 & 29.6 & 2.968 & 710 & $-$0.669 & $-$1.339 & 0.064 & 0.067 \\ 
SZ2 & 0 & 38 & 57.5 & $-$74 & 10 & 0.9 & 2.966 & 364 & $-$0.436 & $-$1.211 & 0.068 & 0.067 \\ 
SZ3 & 0 & 39 & 47.8 & $-$74 & 34 & 44.8 & 2.993 & 134 & $-$0.576 & $-$1.471 & 0.042 & 0.042 \\ 
SZ4 & 0 & 39 & 57.6 & $-$73 & 6 & 3.6 & 2.966 & 978 & $-$0.711 & $-$1.246 & 0.097 & 0.102 \\ 
SZ5 & 0 & 42 & 59.0 & $-$74 & 2 & 44.6 & 2.966 & 541 & $-$0.568 & $-$1.265 & 0.100 & 0.093 \\ 
SZ6 & 0 & 44 & 40.3 & $-$73 & 21 & 51.8 & 2.962 & 1095 & $-$0.676 & $-$1.319 & 0.066 & 0.055 \\ 
SZ8 & 0 & 45 & 16.8 & $-$74 & 42 & 31.1 & 2.937 & 245 & $-$0.657 & $-$1.265 & 0.049 & 0.053 \\ 
SZ9 & 0 & 54 & 23.0 & $-$73 & 31 & 0.2 & 2.974 & 1475 & $-$0.760 & $-$1.161 & 0.074 & 0.083 \\ 
SZ11 & 1 & 0 & 5.7 & $-$71 & 57 & 23.4 & 2.970 & 740 & $-$0.770 & $-$1.278 & 0.070 & 0.068 \\ 
SZ12 & 1 & 0 & 18.3 & $-$74 & 3 & 22.8 & 2.999 & 339 & $-$0.869 & $-$1.127 & 0.048 & 0.090 \\ 
SZ13 & 1 & 1 & 4.7 & $-$73 & 41 & 59.9 & 2.957 & 563 & $-$0.742 & $-$1.306 & 0.082 & 0.076  \\ 
SZ14 & 1 & 2 & 44.9 & $-$72 & 15 & 21.9 & 2.986 & 842 & $-$0.863 & $-$1.244 & 0.034 & 0.073 \\ 
SZ15 & 1 & 5 & 22.5 & $-$71 & 56 & 49.9 & 2.989 & 512 & $-$0.996 & $-$1.197 & 0.058 & 0.067 \\ 
SZ16 & 1 & 7 & 15.6 & $-$74 & 10 & 45.3 & 2.956 & 157 & $-$0.892 & $-$1.266 & 0.062 & 0.063 \\ 
SZ17 & 1 & 7 & 21.6 & $-$72 & 48 & 45.6 & 2.988 & 845 & $-$0.830 & $-$1.144 & 0.040 & 0.032 \\ 
SZ18 & 1 & 8 & 25.4 & $-$73 & 43 & 17.3 & 3.004 & 400 & $-$0.757 & $-$1.339 & 0.064 & 0.060 \\ 
SZ19 & 1 & 8 & 34.8 & $-$71 & 19 & 15.5 & 2.995 & 232 & $-$0.801 & $-$1.208 & 0.073 & 0.092 \\ 
SZ20 & 1 & 11 & 3.0 & $-$72 & 20 & 36.2 & 2.995 & 400 & $-$0.901 & $-$1.387 & 0.056 & 0.051 \\ 
SZ21 & 1 & 14 & 45.3 & $-$71 & 53 & 40.8 & 2.989 & 152 & $-$0.927 & $-$1.239 & 0.068 & 0.077  \\ 
SZ22 & 1 & 15 & 18.7 & $-$73 & 23 & 54.6 & 2.995 & 237 & $-$0.840 & $-$1.145 & 0.058 & 0.054 \\ 
SZ23 & 1 & 15 & 34.1 & $-$72 & 50 & 49.3 & 2.952 & 186 & $-$1.000 & $-$1.185 & 0.071 & 0.076 \\ 
SZ24 & 1 & 17 & 1.0 & $-$71 & 28 & 35.9 & 3.012 & 105 & $-$0.995 & $-$1.264 & 0.077 & 0.080 \\ 
SZ25 & 1 & 18 & 54.5 & $-$74 & 5 & 44.8 & 2.991 & 75 & $-$0.917 & $-$1.145 & 0.057 & 0.066 \\ 
SZ26 & 1 & 20 & 52.4 & $-$72 & 3 & 13.3 & 2.976 & 110 & $-$0.902 & $-$1.255 & 0.056 & 0.054 \\ 
SZ27 & 1 & 20 & 56.1 & $-$73 & 34 & 53.5 & 2.987 & 107 & $-$1.098 & $-$1.196 & 0.064 & 0.049 \\ 
SZ28 & 1 & 21 & 8.4 & $-$73 & 7 & 13.1 & 2.935 & 89 & $-$0.994 & $-$1.103 & 0.103 & 0.093 \\ 
SZ29 & 1 & 24 & 5.8 & $-$72 & 39 & 46.9 & 2.962 & 95 & $-$0.869 & $-$1.153 & 0.110 & 0.099  \\ 
SZ30 & 1 & 26 & 2.7 & $-$73 & 56 & 3.8 & 2.979 & 72 & $-$1.118 & $-$1.307 & 0.100 & 0.106 \\ 
\hline
 & & & \multicolumn{9}{c}{\cite{NK13} PMs} &  \\ \cline{5-11}
\hline
S1 & 0 & 51 & 17.0 & $-$72 & 16 & 51.3 & 1.9 & 42 & $-$0.682 & $-$1.288 & 0.100 & 0.100  \\ 
S2 & 0 & 55 & 34.7 & $-$72 & 28 & 33.9 & 7.6 & 25 & $-$0.722 & $-$1.214 & 0.032 & 0.024 \\ 
S3 & 1 & 2 & 14.5 & $-$73 & 16 & 26.6 & 7.7 & 36 & $-$0.679 & $-$0.974 & 0.026 & 0.028 \\ 
S4 & 0 & 36 & 39.7 & $-$72 & 27 & 42.0 & 2.8 & 10 & $-$0.460 & $-$1.114 & 0.109 & 0.109 \\ 
S5 & 1 & 2 & 34.7 & $-$72 & 54 & 23.8 & 6.8 & 30 & $-$0.806 & $-$1.199 & 0.017 & 0.038 \\ 
\hline
 & & &  \multicolumn{9}{c}{\cite{vdM16} PMs} & \\ \cline{5-11}
\hline
3934 & 0 & 50 & 31.6 & $-$73 & 28 & 42.6 & -- & 1 & $-$0.541 & $-$1.304 & 0.177 & 0.177  \\ 
3945 & 0 & 50 & 38.4 & $-$73 & 28 & 18.1 & -- & 1 & $-$0.668 & $-$1.160 & 0.154 & 0.148 \\ 
4004 & 0 & 51 & 24.6 & $-$72 & 22 & 58.4 & -- & 1 & $-$0.670 & $-$1.165 & 0.148 & 0.143 \\ 
4126 & 0 & 52 & 51.2 & $-$73 & 6 & 53.6 & -- & 1 & $-$0.667 & $-$1.291 & 0.132 & 0.116  \\ 
4153 & 0 & 53 & 4.9 & $-$72 & 38 & 0.2 & -- & 1 & $-$0.821 & $-$1.231 & 0.131 & 0.130  \\ 
4768 & 1 & 1 & 17.0 & $-$72 & 17 & 31.2 & -- & 1 & $-$1.144 & $-$1.239 & 0.151 & 0.143 \\ 
5267 & 1 & 7 & 18.2 & $-$72 & 28 & 3.7 & -- & 1 & $-$0.849 & $-$1.262 & 0.152 & 0.144 \\ 
5714 & 1 & 13 & 30.5 & $-$73 & 20 & 10.3 & -- & 1 & $-$0.992 & $-$1.182 & 0.091 & 0.082 \\ 

\tableline
\end{tabular}
\tablecomments{The identifier used for each data point (the fields or Gaia star ID), and R.A./decl. of reference source (Columns 1, 2, and 3). Column 4 lists the time baseline, in years, between the epochs used to calculate the PM. Column 5 lists the number of stars used in the final transformations after all cuts and iterations have been applied. Columns 6-9 list the observed PMs and errors. } 
\label{tab:pm}
\end{center}
\end{table*}

\section{PM Results} \label{sec:pmresults}
\subsection{Model Design and Analysis} \label{sec:commodel}

Including the PM measurements from NK13 and vdMS16 with our new sample, we have a total of 41 data points for describing the motion of the SMC. This affords us the opportunity to consider the complicating factor of motions internal to the SMC when attempting to determine the COM motion. As can be seen from Figure \ref{fig:reserr}, residual motions are present and significant. 


We fix the radial velocity of the SMC at $v_{\mathrm{sys}}$ = 145.6 $\pm$ 0.6 km s$^{-1}$ \citep{harris06} and the distance modulus at $m - M$ = 18.99 $\pm$ 0.1 \citep{cioni00}. We additionally consider the impact of viewing perspective (projection effects of the 3D COM motion) in the same manner as \cite{vdM02}. While the SMC only subtends $\sim$5 degrees on the sky, this effect can contribute up to $\leq$0.16 mas yr$^{-1}$, a non-negligible fraction of the measured PMs.

For the SMC center, we test two different positions: the H {\small I} kinematical center at ($\alpha$, $\delta$) = ($16.25^{\circ}, -72.42^{\circ}$) \citep{stanimirovic} and the center determined by the structure of the Cepheid population of the SMC at ($\alpha$, $\delta$) = ($12.54^{\circ}, -73.11^{\circ}$) \citep{ripepi17}. With the growing body of work suggesting a disconnect between the stellar motions and the underlying H {\small I} gas, we felt it prudent to examine the new stellar geometric center in addition to the more traditional H {\small I} dynamical center. The locations of both of these centers are shown in Figure \ref{fig:smcareas}. For the centers, we use a fixed uncertainty of 0$^{\circ}$.2 for the position. 

As discussed in Section \ref{sec:introduction}, multiple LOS studies have found evidence for rotation in the SMC. To address this possibility, we test for two different rotation scenarios in our model. For one, we constrain the rotation velocity, $V_{\mathrm{rot}} = 0 \pm$ 40 km s$^{-1}$, and for the second option, we allow $V_{\mathrm{rot}}$ to be a free parameter. For both, we treat the rotation as rising out to a radius of 0.6 kpc and then constant after that. As a fifth case, we also test for allowing both $V_{\mathrm{rot}}$ and the center position to be free fit parameters. However, we find that the data are unable to provide a useful constraint on the center position, converging to a center close to the H {\small I} center but with an uncertainty of 4 degrees.

Most of the LOS studies focused on the innermost 2 degrees of the SMC, an area that we do not sample well. Instead, most of our statistical leverage comes from the outer regions of the SMC. As we are limited in our sampling density, we opt to keep the inclination of the model near 0$^{\circ}$. In total, we have four cases to test, two choices of the SMC center and two options for its internal rotation about each center.

For the COM PM itself, we leave it as a free parameter, optimized by minimizing the model's $\chi ^{2}$ with respect to the data. The fit statistic is the same as that used by \cite{vdM14},

\begin{equation} 
\begin{split}
\chi_{\mathrm{PM}}^2 \equiv \sum_{i=1}^{M} [(\mu_{W,\mathrm{obs},i} - \mu_{W,\mathrm{mod},i})/\Delta\mu_{W,\mathrm{obs},i}]^2 \\
+ [(\mu_{N,\mathrm{obs},i} - \mu_{N,\mathrm{mod},i})/\Delta\mu_{N,\mathrm{obs},i}]^2,
\end{split}
\label{eq:chi2}
\end{equation}
The resulting parameters for the minimized $\chi ^{2}$ model are used to create mock data, using a Monte Carlo approach. As in NK13, these mock data are used to estimate the uncertainties in the best fit parameters for the model. Each set of mock data are given uncertainties, drawn from the the observational uncertainties but scaled by a factor of ($\chi_{\mathrm{min}}^{2}$/$N_{\mathrm{dof}}$)$^{1/2}$ to compensate for any underestimate of the uncertainties, where $N_{\mathrm{dof}} = N_{\mathrm{data}} - N_{\mathrm{param}} + N_{\mathrm{fixed}}$, and $\chi_{\mathrm{min}}^{2}$ is the minimum fit statistic for the model. We generate and fit multiple sets of mock data, and then we use the dispersion in the parameters found as an estimate of the random uncertainty.




 
\subsection{COM Results} \label{sec:compm}

\begin{table*}[t]
\begin{center}
\caption{SMC Best Fit Parameters.}
\begin{tabular}{ccccccc}
\tableline
\tableline
(1) & Center & & H {\small I} & H {\small I} & R17 & R17 \\
(2) & $V_{\mathrm{rot}}$ & & Constrained & Free & Constrained & Free\\
\tableline
(3) & $\mu_{W}$ & mas yr$^{-1}$ &  $-$0.80 $\pm$ 0.11 & $-$0.83 $\pm$ 0.02 & $-$0.74 $\pm$ 0.03 & $-$0.73 $\pm$ 0.02 \\ 
(4) & $\mu_{N}$ & mas yr$^{-1}$ &  $-$1.21 $\pm$ 0.04 & $-$1.21 $\pm$ 0.01 & $-$1.25 $\pm$ 0.13 & $-$1.24 $\pm$ 0.02 \\ 
(5) & $\mu_{\mathrm{tot}}$ & mas yr$^{-1}$ & \multicolumn{1}{r}{1.45 $\pm$ 0.12 } & \multicolumn{1}{r}{1.47 $\pm$ 0.02} & \multicolumn{1}{r}{1.45 $\pm$ 0.13} & \multicolumn{1}{r}{1.45 $\pm$ 0.03} \\ 
(6) & $V_{\mathrm{rot}}$ & km s$^{-1}$ & --- & $-$11.6 $\pm$ 4.0 & --- & $-$0.3 $\pm$ 3.5 \\ 
(7) & ($\chi_{\mathrm{min}}^{2}$/$N_{\mathrm{dof}}$)$^{1/2}$ & & 2.32 & 2.29 & 2.32 & 2.37 \\ 

\hline

\tableline
\end{tabular}
\tablecomments{Line 1 indicates the center used in the model as described in Section \ref{sec:commodel}, and line 2 indicates whether V$_{\mathrm{rot}}$ was left free or constrained to 0 $\pm$ 40 km s$^{-1}$. Lines 3-6 are the best fit values for $\mu_{W}$, $\mu_{N}$, $\mu_{\mathrm{tot}}$, and V$_{\mathrm{rot}}$, respectively, for the four models. The units for each parameter are listed in the adjacent column. Line 7 is the statistic used to assess the quality of fit of the model to the data, described in Section \ref{sec:commodel}. R17 refers to the estimate of the center of the SMC from \cite{ripepi17}. \\
$^{a}$ Refers to \cite{ripepi17}\\}
\label{tab:bestfit}
\end{center}
\end{table*}

The final best fit parameters for each of the four models are listed in Table \ref{tab:bestfit}. We see that the choice of the dynamical center does have an effect on the estimated COM PMs, differing by $\sim$3$\sigma$ in $\mu_{W}$ and by $\sim$2$\sigma$ in $\mu_{N}$. To reflect this uncertainty in the COM motion, we add a systematic error term to our final PM measurement, which we define as the difference between the best fit PM values for the $V_{\mathrm{rot}}$ free cases (see the discussion below). For $\mu_{W}$ this is 0.1 mas yr$^{-1}$, and for $\mu_{N}$ is 0.03 mas yr$^{-1}$. Additionally, the choice of the center seems to affect the likelihood of a detection of a rotation signature when $V_{\mathrm{rot}}$ is allowed to be free. The H {\small I} center converges on $V_{\mathrm{rot}} = 12 \pm$4 km s$^{-1}$ (random error only), while the geometrically determined center is consistent with no rotation, $V_{\mathrm{rot}} = 0 \pm $4 km s$^{-1}$ (random error only). When we consider the impact of the systematic error term, both rotation signatures become statistically consistent with no rotation.
While the model using the H {\small I} center and allowing $V_{\mathrm{rot}}$ to be free does formally produce the best fit, the differences are not significant (see Table \ref{tab:bestfit}). This underscores the difficulty of using a simple model to describe the potentially complex nature of the SMC internal kinematics. 


We choose the $V_{\mathrm{rot}}$ free, H {\small I} center model for our final estimate and comparison with previous studies (seen in Table \ref{tab:prevcom}) because it formally has the smallest $\chi_{\mathrm{min}}^{2}$/$N_{\mathrm{dof}}$ and most previous works have adopted the H {\small I} dynamical center.\footnote{We do examine the impact of this choice of dynamical center on our subsequent orbital modeling.} All four of our new COM motion estimates are statistically consistent with the prior values found for the SMC given the uncertainties. 
These is a slight offset between our work and vdMS16 as compared to NK13 and \cite{cioni16}. In the latter two studies, the majority of the measurements come from the western half of the SMC, while the first two have more uniform coverage of the whole SMC. This underscores the complex nature of the SMC and the care that must be taken to avoid contamination of the global PM estimate by local motions. As a consistency check, we also consider the 28 new HST fields by themselves. For the two choices of SMC center, and a fixed versus free rotation signal, we find results that agree within the random errors of the full sample. This is perhaps not surprising given that these 28 fields account for the majority of the 41 total measurements considered here.

For the TGAS PM errors, a systematic effect that is not explicitly included is possible spatial correlations in the PM errors \citep{lindegren16}. The effect of such correlations would be to underestimate the random error in the weighted average PM of the sample. However, the agreement between the TGAS and HST results shows that any residual systematic errors must be below the random errors. Similarly for the HST data, the main possible residual systematic errors are from the geometric distortion solution and charge transfer efficiency effects. Both are expected from \cite{bellini11} and \cite{anderson14} to be below our random errors. The main systematic uncertainty, which is larger than our random errors, comes from not being able to establish a dynamical center for the SMC from our data alone.

\begin{table}[t] 
\begin{center}
\caption{SMC COM PMs.}
\begin{tabular}{cccc}
\tableline
\tableline
Work & Data & $\mu_{W}$ & $\mu_{N}$ \\
 & & (mas yr$^{-1}$) & (mas yr$^{-1}$) \\
\tableline
\textbf{This paper}  & \textit{HST}+\textit{Gaia} & $-$0.83$\pm$0.02 & $-$1.21$\pm$0.01\\ 
vdMS16 & \textit{Gaia} & $-$0.87$\pm$0.07 & $-$1.23$\pm$0.05  \\ 
\cite{cioni16} &  VMC  & $-$0.81$\pm$0.07 & $-$1.16$\pm$0.07 \\ 
NK13 & \textit{HST} & $-$0.77$\pm$0.06 & $-$1.12$\pm$0.06 \\ 
\cite{vieira10} & SPM$^{a}$ & $-$0.98$\pm$0.30 & $-$1.10$\pm$0.29 \\ 

\hline

\tableline
\end{tabular}
\tablecomments{Column 1 indicates the source of the measurement and Column 2 the type of data used to determine the result. \\
$^{a}$ Yale/San Juan Southern Proper Motion program \\}



\label{tab:prevcom}
\end{center}
\end{table}

\section{Internal Kinematics} \label{sec:internal}
\subsection{Full Star Sample} \label{sec:fullstar}

\begin{figure*}
\begin{center}
 \includegraphics[width=5.6in]{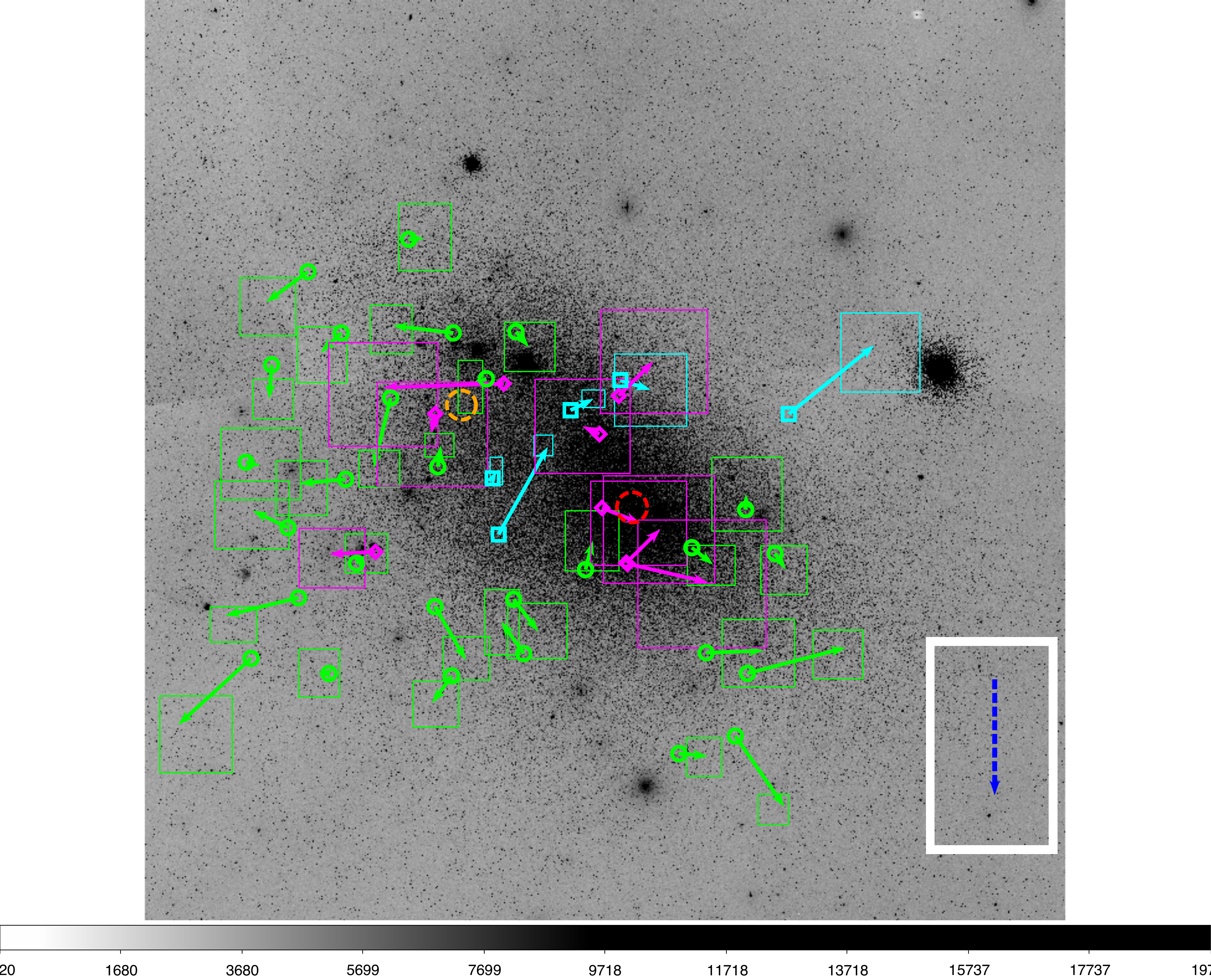} 
 \caption{The residual motion vectors calculated after subtracting our adopted COM motion determined in this study. Similar to Figure \ref{fig:smcpos} the green, cyan, and magenta measurements are this study, NK13, and vdMS16, respectively. The boxes indicate the uncertainty in the motion for that field where a vector that exceeds its box corresponds to a residual vector of greater than $1 \sigma$. A reference vector of 100 km s$^{-1}$ is shown at the bottom right, and the two centers used for the models are shown as well with the H {\small I} derived center marked by the dashed orange circle and the \cite{ripepi17} center marked by the dashed red circle. For orientation, north is up and east is to the right. }
   \label{fig:reserr}
\end{center}
\end{figure*}

We can now subtract the global COM PM, including the perspective motion, to find the internal motions of the SMC. The result is shown in Figure \ref{fig:reserr}. In addition to the calculated vectors, the observational error is also shown so that the significance of a particular vector can be evaluated.

\begin{figure}
\begin{center}
 \includegraphics[width=3.4in]{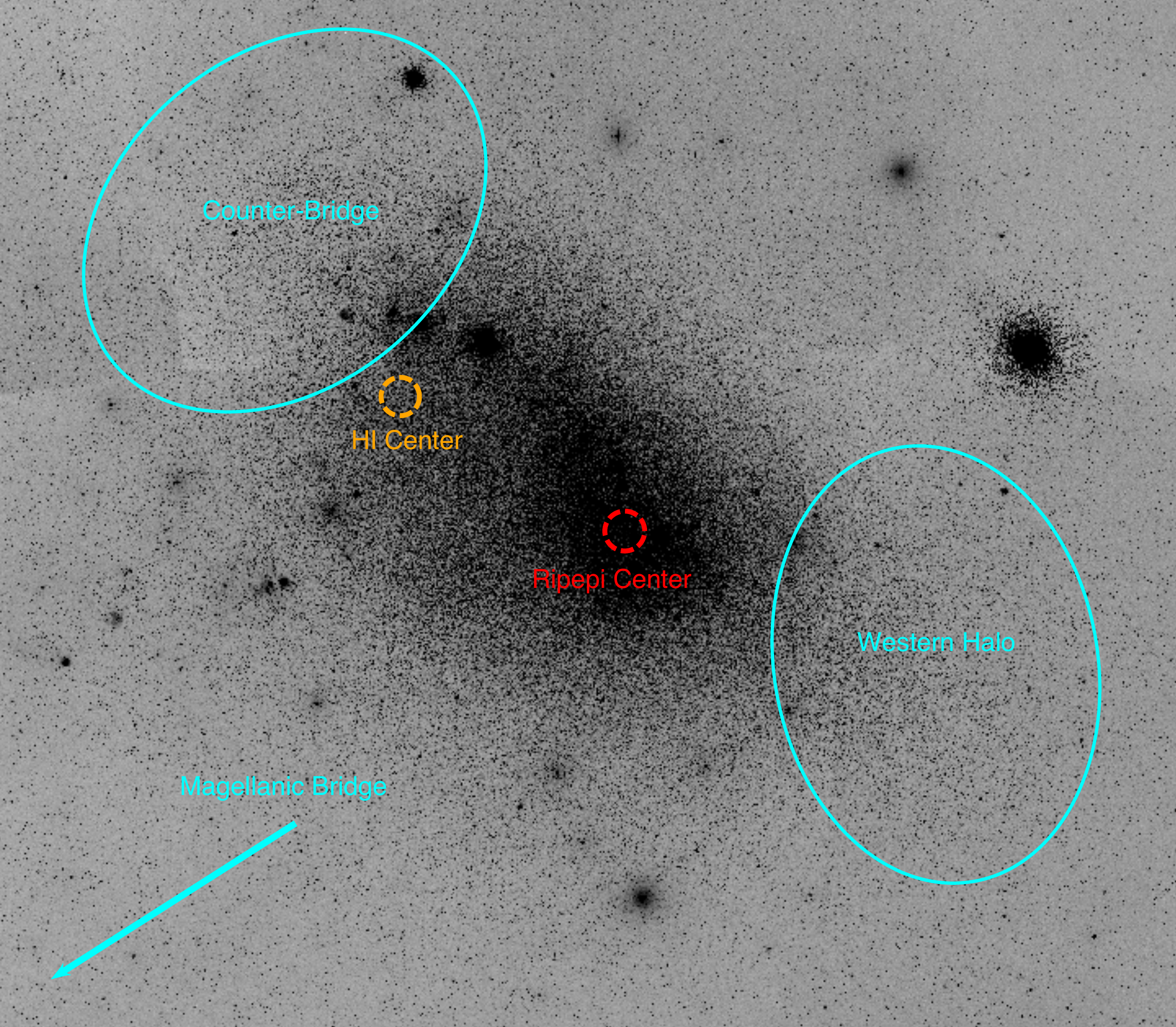} 
 \caption{The locations of the two centers tested: the H {\small I} derived center (dashed orange circle) and the \cite{ripepi17} center (dashed red circle). Three areas with potential kinematic signatures are also marked. The regions considered to be the Counter Bridge and Western Halo are marked by the cyan ellipses, and the direction toward the Magellanic Bridge is marked by the cyan vector. For orientation, north is up and east is to the right.}
   \label{fig:smcareas}
\end{center}
\end{figure}

\begin{figure}
\begin{center}
 \includegraphics[width=3.4in]{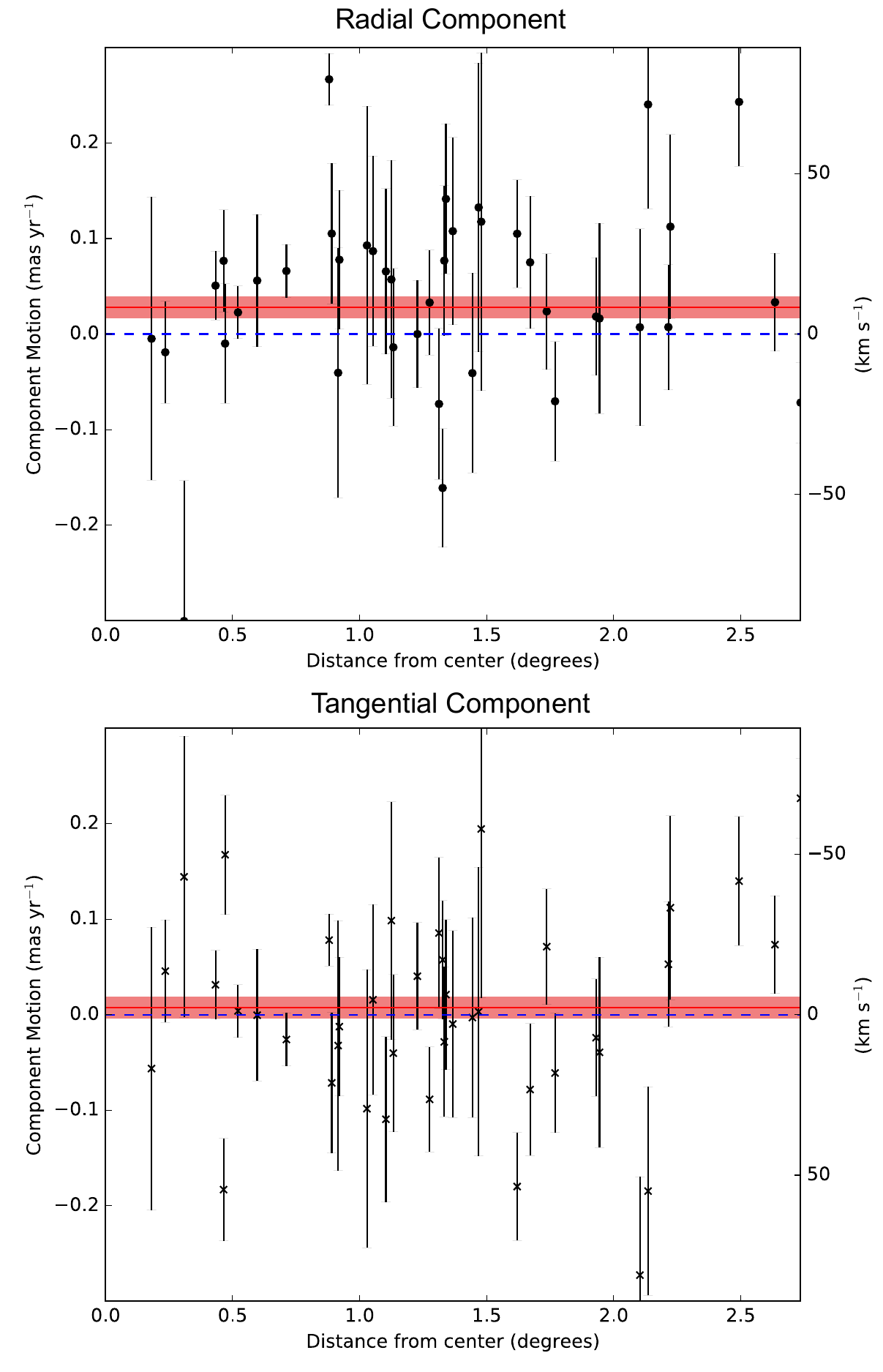} 
 \caption{Amplitudes for the radial (top) and tangential (bottom) components of the residual motions for the H {\small I} center, V$_{\mathrm{rot}}$ free model, as a function of distance from the center. For calculating the error-weighted mean, we exclude any fields not consistent with zero to within twice the observational error. The resulting error-weighted mean is indicated by the red line and the error of the weighted mean is shaded above and below the line.}
   \label{fig:comps}
\end{center}
\end{figure}

At first glance, there does not appear to be any coherent rotational structure to the residual vectors. As an alternative visualization, we decompose each field into its radial and tangential components, $\mu_{\mathrm{res,rad}}$ and $\mu_{\mathrm{res,tan}}$ respectively, as a function of distance from the center, and calculate the error-weighted mean and the error for the weighted mean, shown in Figure \ref{fig:comps}. In calculating the error-weighted mean, we exclude any fields not consistent with zero to within twice the observational error. For the radial component, we find for the H {\small I} center $\bar{\mu}_{\mathrm{res,rad}}$ = 0.027 $\pm$ 0.010 mas yr$^{-1}$ and for the \cite{ripepi17} center $\bar{\mu}_{\mathrm{res,rad}}$ = 0.015 $\pm$ 0.010 mas yr$^{-1}$. For both centers, a radial motion greater than zero is preferred, consistent with a tidally disrupting system. For the tangential component, we find for the H {\small I} center $\bar{\mu}_{\mathrm{res,tan}}$ = 0.008 $\pm$ 0.010 mas yr$^{-1}$ and for the \cite{ripepi17} center $\bar{\mu}_{\mathrm{res,tan}}$ = 0.001 $\pm$ 0.010 mas yr$^{-1}$. If a rotation signal were present, the fields would be offset from zero, but both means are consistent with zero.

In the southwest and southeastern regions, large and statistically significant residuals can be seen. For the southeastern region, this coincides with the direction toward the Magellanic Bridge (shown in Figure \ref{fig:smcareas}), peaking around 80 km s$^{-1}$. This is the first measured stellar motion away from the SMC and toward the Bridge. In the southwestern region, the strong coherent motions appear to be coincident with the ``Western Halo,'' identified by \cite{dias16}. The other potential dynamic signature, the  ``Counter Bridge'' (e.g., \citealt{besla11}), which is predicted in the northeastern section of the SMC, does not appear as a prominent feature in our data. Either the way, the general finding of ordered mean motion radially away from the SMC in the outer regions of the galaxy, provides kinematic evidence that the SMC is in the process of tidal disruption.

The combination of the amplitude of these residual vectors and their spatial coherence suggest the possibility of some of these fields being unbound from the SMC. To provide a physical sense for what might be unbound, we estimate the escape speed. We relate the escape speed $v_{\mathrm{e}}$ to the circular velocity $v_{\mathrm{c}}$ under the simple assumption of a Kepler potential, for which 
\begin{equation}
\frac{v_{\mathrm{e}}^2}{2} = \frac{GM}{R}\ , v_{\mathrm{c}}^2 = \frac{GM}{R},
\end{equation}
so that $v_{\mathrm{e}} = \sqrt[]{2} v_{\mathrm{c}}$. For $v_{\mathrm{c}}$, we use results from \cite{vdmf93} to relate it to the LOS velocity dispersion $\sigma_{\mathrm{LOS}}$. If we assume an isotropic velocity distribution and a density profile of $r^{-3}$, from Eq. (B6b) in \cite{vdmf93} we find $\sigma_{\mathrm{LOS}} = \sqrt[]{\pi/16}\ v_{\mathrm{c}}$. Combining these two relations together, we get $v_{\mathrm{e}} = 3.19\ \sigma_{\mathrm{LOS}}$. Using the measurement from \cite{dobbie14a} for $\sigma_{\mathrm{LOS}} \approx 26$ km s$^{-1}$, we find a final $v_{\mathrm{e}} \approx 83$ km s$^{-1}$. We note that this is a lower limit, since realistic potentials are more extended than a Kepler potential, but it provides a useful intuition for the state of the SMC.
Several of our fields have a total residual motion consistent with this estimate of  the escape velocity. This provides kinematical evidence that the stars there could be unbound. This is consistent with the fact that other material from the SMC that is now at larger radii than the radii where we are probing, must have become unbound to form the Magellanic Stream and Magellanic Bridge.

We wanted to examine the impact on the COM PM from narrowing our choice of fields included. For stars in equilibrium around the COM, one expects to measure a radial PM residual of zero, calculated as in Figure \ref{fig:comps}. So we discard all
fields for which the residual is not consistent with zero to within twice the observational error. This excludes five fields, 2 from our new sample, 1 from vdMS16, and 2 from NK13. After fitting our model to this restricted  subsample, the resulting COM PMs do not significantly vary from the original values for their respective centers. The choice of center has a bigger effect on our data than this difference in field selection.



\subsection{Red versus Blue Stellar Motion} \label{sec:redblue}


With multiple LOS studies in potential tension over the behavior of different stellar populations in the SMC, as discussed in Section \ref{sec:introduction}, we wanted to explore our data's ability to constrain this problem. 

\begin{figure}
\begin{center}
 \includegraphics[width=3.4in]{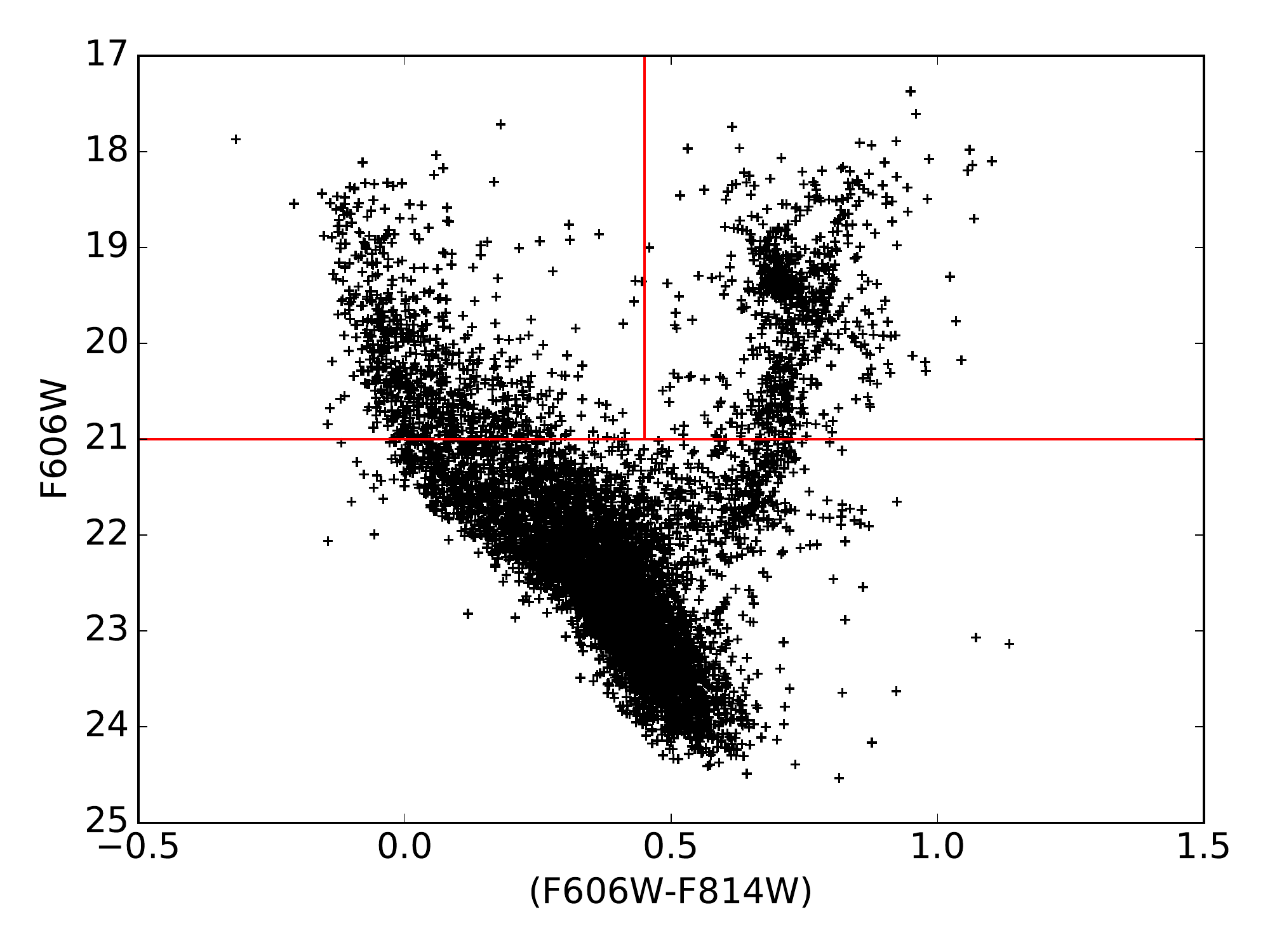} 
 \caption{The composite CMD for all new SMC fields. The dividing lines for the red and blue subpopulations, examined in Figure \ref{fig:rgbwcs}, are marked at (F606W$-$F814W)=0.45 and F606W$<$21 mag.}
   \label{fig:smccmd}
\end{center}
\end{figure}

We selected samples of red and blue stars, separated by a color of (F606W $-$ F814W) = 0.45 and with F606W $<$ 21 mag, as shown in Figure \ref{fig:smccmd}. This cleanly delineates the two populations, and we will refer to these as the red and blue populations. For a PM to be calculated, we also require that the field has a minimum of 10 stars in each subsample.

For each population of stars, we repeated the process of iteratively transforming the source positions into a master frame, as described in Section \ref{sec:twoepoch}. Due to the smaller number of fields with enough stars, we calculated a simple weighted average for the systemic motion of the fields and use that, along with the contributions from SMC geometry (viewing perspective) calculated from the model, to create residual motions as a function of color. For the red population, the resulting systemic motion is $\mu_{W}$ = $-$0.72 $\pm$ 0.06 mas yr$^{-1}$ and $\mu_{N}$ = $-$1.23 $\pm$ 0.08 mas yr$^{-1}$, while the blue population was found to have an average motion of $\mu_{W}$ = $-$0.81 $\pm$ 0.06 mas yr$^{-1}$ and $\mu_{N}$ = $-$1.24 $\pm$ 0.08 mas yr$^{-1}$.
These estimates are statistically consistent but the differences may be real.
The fields with a large enough number of red stars tend to lie toward the southwestern portion of the SMC, while the fields that have enough blue stars lie toward the eastern side of the SMC. More western fields will bias the average motion toward a greater western motion (a smaller $\mu_{W}$) while more fields near the Bridge 
will bias it in the opposite direction.

\begin{figure}
\begin{center}
 \includegraphics[width=3.4in]{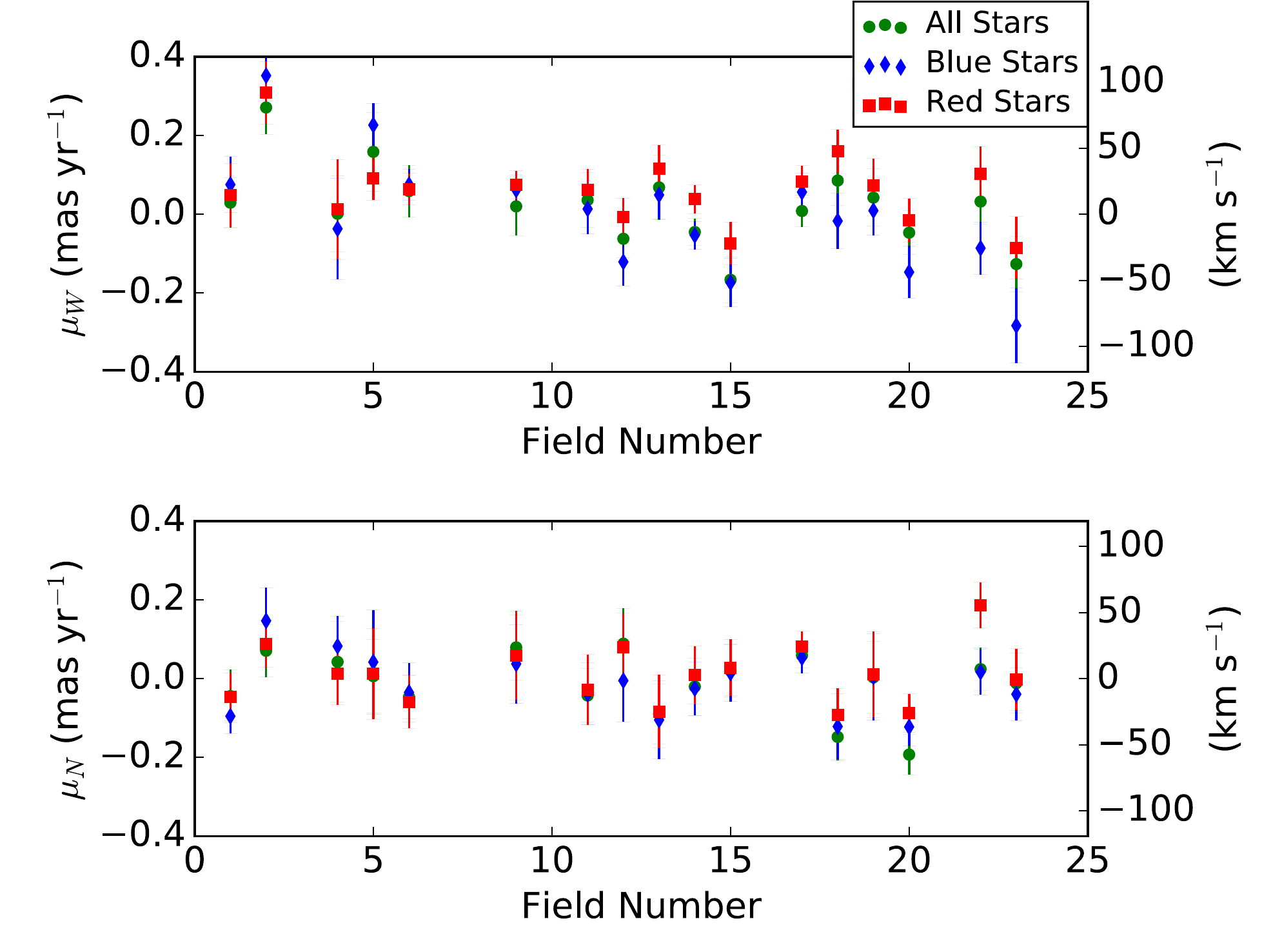} 
 \caption{Residual PMs in the E$-$W (top) and N$-$S (bottom) directions for all fields with enough red and blue stars for the transformations. There are essentially no statistically significant differences between the different populations.}
   \label{fig:rgbwcs}
\end{center}
\end{figure}

Indeed, when we examine fields that have both red and blue stars we see that for many of the fields there are no significant differences between the motions of the populations (see Figure \ref{fig:rgbwcs}). Only for the highest field numbers, corresponding to fields on the outer edges of the SMC, do we note a significant difference. Unfortunately, those fields are also among the sparsest, often falling on the threshold of the 10 required stars. 
While the ability of WFC3/UVIS to detect enough stars to be able to distinguish between different stellar populations in the SMC is exciting, these results suggest it will require much better coverage of the entire SMC, rather than the pencil beam investigation undertaken here, to discern a difference between the populations, if one exists.

\section{Orbit Implications} \label{sec:orbit}

The past orbits of the LMC$-$SMC system about the MW have been explored in many previous works. These have tended to concentrate on the more massive member of the pair, for instance, evaluating whether the LMC is on a first passage (\citealt{besla07}; 
NK13), exploring whether the presence of the LMC influences the dynamics of other MW satellites or even the MW-LMC barycenter or MW disk (\citealt{veraciro13}; 
\citealt{gomez15}; 
\citealt{laporte17, laporte18}; 
\citealt{patel17}), 
and exploring substructure that might have come in with the LMC (
\citealt{yozin15}; 
\citealt{jethwa16a}; 
\citealt{sales17}; 
\citealt{fritz17, fritz18}; 
\citealt{sales11}). 
However, the relationship of the Clouds to each other, specifically how long they have been a binary, and how closely they may have interacted in the past, is still relatively unexplored and unconstrained (e.g., 
\citealt{besla12}; 
\citealt{diaz12}). 
Our new data afford us a much better constraint on the past interactions between the Clouds themselves. In particular, the impact parameter and timing of the last encounter between the Clouds is critical to our understanding of the formation of the Magellanic Bridge and the internal structure of both galaxies (e.g., \citealt{gardiner96}; \citealt{yoshizawa03}; \citealt{bekki07}; \citealt{ruzicka10}; \citealt{besla16}; \citealt{diaz12}; \citealt{guglielmo14}; \citealt{pardy18}), and this is where we focus our modeling efforts.

\subsection{Methodology} \label{subsec:orbitmethod}
Our orbital modeling procedure is basically identical to that in NK13, and we refer the interested reader to that work for the particulars. The MW is modeled is an axisymmetric three-component potential with a Navarro$-$Frenk$-$White (NFW) halo 
\citep{nfw96,nfw97}
, Miyamoto$-$Nagai disk 
\citep{miyamoto75}, 
and a Hernquist bulge 
\citep{hernquist90}
. The NFW halo is adiabatically contracted to account for the presence of the disk 
\citep{gnedin04}
, and the NFW density profile is also truncated at the virial radius. We explore two such MW models that span the mass range of recent studies: a light model with a total virial mass of $1\times 10^{12}M_{\odot}$, and a heavy model with a mass of $2 \times 10^{12}M_{\odot}$ (e.g., 
\citealt{blandhawthorn16}
). As we saw in NK13, a high-mass MW tends to disrupt the LMC$-$SMC binary in the past, while it is easier for them to have been bound for longer in a low-mass MW model.

Our LMC model is slightly different than that used in NK13, but still spans a low- and high-mass range. It is less likely that the LMC and SMC have been a long-lived binary if the LMC mass is low, and much more likely if the LMC mass is high. Our low-mass LMC model of $3.7 \times 10^{10} M_{\odot}$ comes from requiring the rotation curve to be flat at a value of 91.7 km s$^{-1}$ 
\citep{vdM14} 
out to 20 kpc. To make sure that the adopted mass profile matches the dynamical mass of $1.3 \times 10^{10} M_{\odot}$ at 9 kpc 
\citep{vdM09}
, the LMC is modeled as a Plummer potential with a softening length of 9 kpc. Our high-mass LMC of $1.8 \times 10^{11} M _{\odot}$ is motivated by the minimum LMC mass that allows the LMC and SMC to have been a long-lived binary even in the presence of a massive MW (NK13) and cosmological expectations \citep{moster13}. Here, the LMC is also modeled as a Plummer potential with a softening parameter of 20 kpc. As in NK13 the SMC mass is assumed to have been tidally truncated by the LMC at early times, and its mass is kept fixed at $3 \times 10^9 M_{\odot}$ 
\citep{stanimirovic}
.

We draw 10,000 random values for the LMC and SMC PM (NK 13 and this work, respectively), distances (\citealt{cioni00}; \citealt{freedman01}), 
 and line-of-sight velocities (\citealt{vdM02}
; 
\citealt{harris06}
). The Galactocentric distances and velocities are calculated using the same conventions as used in NK13. Since the LMC PM is the same as in that work, we also use the same solar parameters for consistency 
\citep{mcmillan11}
. These values are broadly consistent with other studies such as that of 
\cite{bovy12}
. This Monte Carlo method allows us to properly take into account any covariances in the uncertainties of the measured parameters of the Clouds and the Sun. The resulting mean values for the present-day Galactocentric velocity and relative velocity are shown in Table \ref{tab:galacto}. We then follow the orbits of the LMC and SMC backward in time for the four combinations of LMC and MW mass models.


\begin{table*}[t]
\begin{center}
\caption{Galactocentric Velocities}
\begin{tabular}{cccrrrrrr}
\tableline
\tableline
Galaxy & PM & & $v_{X}$ & $v_{Y}$ & $v_{Z}$ & $v_{\mathrm{tot}}$ & $v_{\mathrm{rad}}$ & $v_{\mathrm{tan}}$\\
& & & (km s$^{-1}$) & (km s$^{-1}$) & (km s$^{-1}$) & (km s$^{-1}$) & (km s$^{-1}$) & (km s$^{-1}$) \\
\tableline

SMC & This Work & & 18 $\pm$ 6 & $-$179 $\pm$ 16 & 174 $\pm$ 13 & 250 $\pm$ 20 & $-$10 $\pm$ 1 & 250 $\pm$ 20\\ 
LMC & Three-epoch NK13 &  & $-$57 $\pm$ 13 & $-$226 $\pm$ 15 & 221 $\pm$ 19 & 321 $\pm$ 24 & 64 $\pm$ 7 & 314 $\pm$ 24 \\ 
SMC-LMC & ... &  & 75 $\pm$ 17 & 47 $\pm$ 22 & $-$47 $\pm$ 23 & 103 $\pm$ 26 & 92 $\pm$ 29 & 43 $\pm$ 11 \\ 

\hline

\tableline
\end{tabular}
\tablecomments{The three lines list the SMC velocity, the LMC velocity, and the relative velocity between the SMC and LMC, as measured in this work. Column 1 lists the galaxy name. Column 2 lists the assumed PM value, where the SMC value is this work's value for the COM PM estimate, assuming the H {\small I} center and fitting for $V_{\mathrm{rot}}$, and the LMC value is taken from NK13. To correct for the solar reflex motion, we use the improved \cite{mcmillan11} value of $V_{0}$ = 239 $\pm$ 5 km s$^{-1}$ and the improved \cite{schonrich10} solar peculiar velocity. Columns 3$-$5 list the Galactocentric velocity coordinates ($v_{X}$, $v_{Y}$, $v_{Z}$). Columns 6$-$8 list the total length of the velocity vector, the radial component, and the transverse component, respectively. Uncertainties were calculated in a Monte Carlo fashion that propagates all relevant uncertainties in the position and velocity of both the Clouds and the Sun. Distance uncertainties are based on $\Delta m - M$ = 0.1. Velocity uncertainties in the Galactocentric frame are highly correlated, because uncertainties in the LOS direction than in the transverse direction.   \\}
\label{tab:galacto}
\end{center}
\end{table*}

\subsection{Impact parameter and timing of the last SMC-LMC encounter} \label{subsec:lmc}
\begin{figure*}
\begin{center}
 \includegraphics[width=6.8 in]{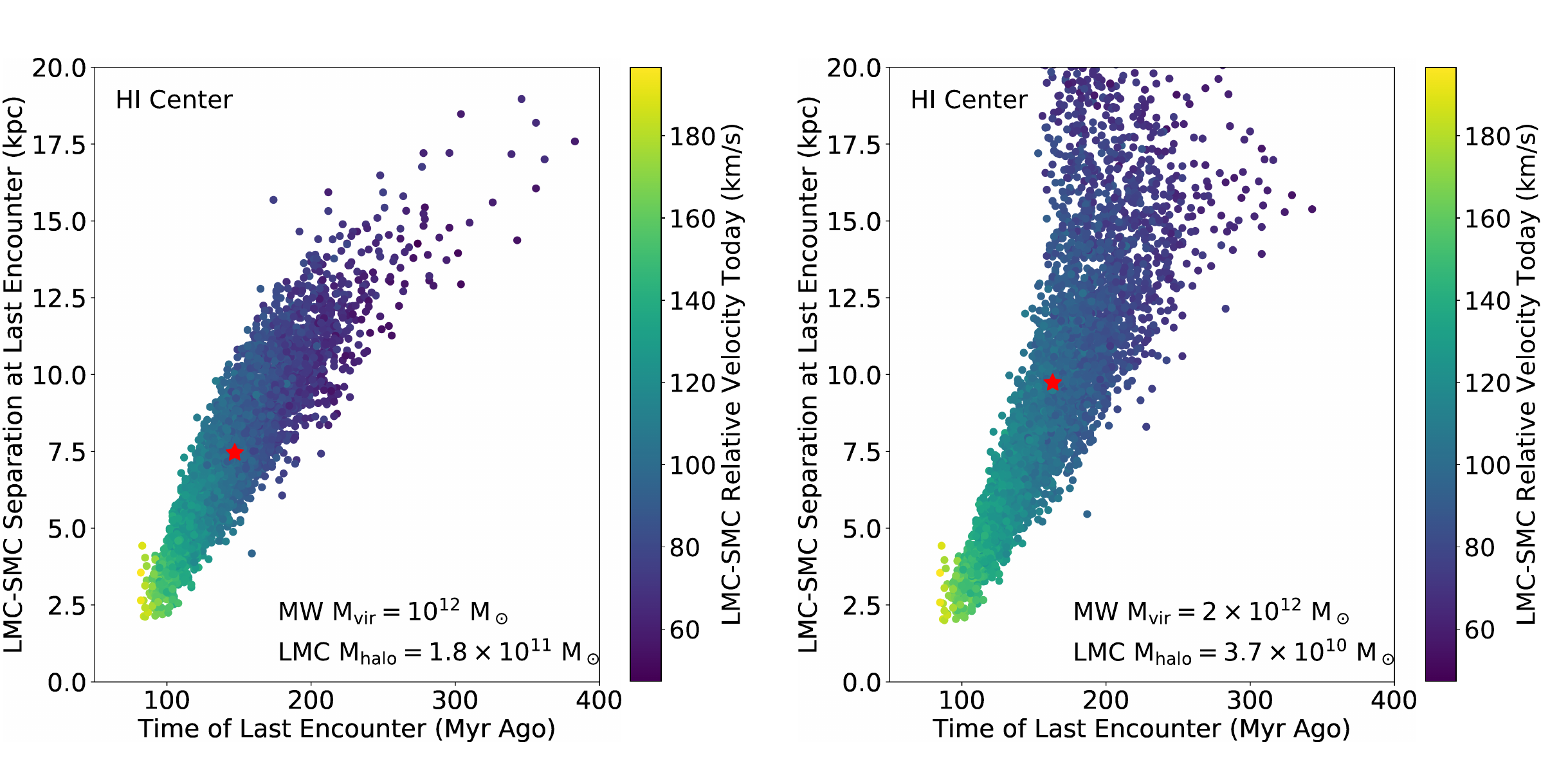} 
 \caption{The LMC$-$SMC separation at their last encounter as a function of the time of the last encounter, color coded by the present-day relative velocities between the LMC and SMC (in km s$^{-1}$). The panel on the left shows results for integrations assuming a MW virial mass of 10$^{12}$ M$_{\odot}$ and an LMC halo mass of 1.8$\times$10$^{11}$ M$_{\odot}$, on the larger (lower) end of MW (LMC) mass possibilities in the literature. The panel on the right is for a heavier MW virial mass of 2$\times$10$^{12}$ M$_{\odot}$ and a lighter LMC halo mass of 3.7$\times$10$^{10}$ M$_{\odot}$, on the larger(lower) end of possible MW(LMC) masses. The red star indicates the mean value in each case. Impact parameters higher than 20 kpc are found to be highly unlikely. Impact parameters as small as 2.5 kpc and as recent as 100 Myr are possible, but a bullseye hit (0 kpc impact parameter) is unlikely.}
   \label{fig:HIorbit}
\end{center}
\end{figure*}

We are interested to see if we can constrain the likelihood of a past collision between the Clouds. We therefore keep track of the minimum separation achieved between the Clouds in the past, and the time of that ``encounter.'' As expected, the extremes of the possible distributions in this encounter come from a low-mass LMC with a high-mass MW, and a high-mass LMC with a low-mass MW. We therefore only show the outcomes for these two mass combinations in Figure \ref{fig:HIorbit}. We find that the choice of SMC center makes no discernible difference -- the minimum separations and encounter times agree to within the errors -- and so we show the results for the H {\small I} center only.

In Figure \ref{fig:HIorbit}, we show the LMC$-$SMC separation at the last close encounter (in kpc), versus the time of the last encounter (Myr in the past), color coded by the present-day relative velocity between the Clouds (km s$^{-1}$). For both mass combinations, there is a strong trend with relative velocity, with the highest relative velocities resulting in the most recent and closest encounters. The fact that the highest relative velocities result in the most recent past encounters makes more obvious sense, but the reason that they also result in the closest encounters between the Clouds is because the highest relative velocities correspond to the largest angles between the LMC and SMC velocity vectors today. 

The majority of cases (97\%) result in a past encounter in which the centers of the Clouds come closer to each other than 20 kpc, which is the currently observed northern extent of the LMC disk \citep{mackey16}. This result is robust to choices in MW and LMC mass. 
Specifically, the minimum separation between the centers of mass of the two Clouds has a mean value of $7.5 \pm 2.5$ kpc about $147 \pm 33$ Myr ago in the case of the heavy LMC and light MW model, and $9.7 \pm 4.5$ kpc about $163 \pm 36$ Myr ago in the case of the light LMC and heavy MW model. In both of the considered mass cases, the smallest separation achieved is $\sim$2 kpc, indicating that the Clouds could have experienced a direct collision, 
 but a bullseye hit (0 kpc impact parameter) is unlikely. Also, only 3\% of cases have minimum separations larger than 20 kpc, for all mass combinations considered.
 
It seems extremely likely that the Clouds have hit each other (since the LMC's disk radius is 18.5 kpc; \citealt{mackey16}). The fact that the SMC is in reality an extended body, and here we plot only the separation of the COMs, strengthens this argument further. This result is consistent with the model of \citet{besla12}, specifically their Model 2, in which the LMC and SMC pair have experienced a recent direct collision roughly 100 Myr ago, which also produces the off-center stellar bar and one-armed spiral of the LMC (see also \citealt{bekki07}; \citealt{pardy16}).

In previous work the timing of the last encounter has been estimated at $< 300$ Myr based on the age of stellar populations in the Bridge \citep{harris06}. The previous PMs \citep{NK06a} were consistent with a timing of $\sim 150$ Myr ago \citep[see][]{ruzicka10}, but now the error bars on this PM estimate are lower, supporting a very recent encounter. Interestingly, if you make the simple assumption that the radial expansion velocity is comparable to the tangential signal ($\sim$80 km s$^{-1}$), in the 150 Myr since the SMC would have expanded roughly 12 kpc. With the deep LOS extension in the eastern region of $\sim$23 kpc \citep{nidever13} and assuming an intrinsic size for the SMC of $\sim$10 kpc (found in the western region by \citealt{mackey18}), this leaves an unaccounted for expansion of $\sim$ 13 kpc, which roughly coincides with our preferred timing and expansion velocity. Note that this encounter is still before the pericentric approach of the LMC to the MW (which happened $\sim 50$ Myr ago). Also previous works typically take the impact parameter to be around 10 kpc or larger \citep{ruzicka10,diaz12}. 
Now we have both refined the impact parameter to be smaller as well as ruled out larger impact parameters, supporting a collision model.

\section{Discussion and Conclusions} \label{sec:disc}

We have analyzed two epochs of PM data for 30 new fields in the SMC with \textit{HST} WFC3/UVIS. We combine these data with previous \textit{HST} PM results from NK13 and \textit{Gaia} PM results from vdMS16 to create the largest PM data set yet for the SMC. Here we summarize our results, the new implications for the SMC's history with the LMC, and future directions for the work.

\subsection{Conclusions for PMs} \label{subsec:pmconc}

With the $\sim$3 year baseline, our analysis gives results that have a range of errors comparable to the errors from NK13. In fields with a large number of stars in the final transformation, we find errors of order $\sim$0.03 mas yr$^{-1}$, similar to those for the $\sim$7 year baseline measurements from NK13. Where our fields become sparser, our errors increase toward a maximum comparable to the errors from NK13, $\sim$0.1 mas yr$^{-1}$, that also had relatively sparse star fields and a shorter baseline.

We have successful measurements for 28 of the 30 fields (as explained in Section \ref{sec:data}), and we combine this data set with the PMs from NK13 and vdMS16 to improve our sampling of the SMC and more tightly constrain the estimate for the SMC COM PM. We fit this data set to a model for the SMC similar to the one laid out in \cite{vdM02}, leaving only the PM of the SMC and a possible rotational velocity as free parameters. We find that our dataset by itself does not allow us to independently determine the dynamical center of the SMC to better precision than previous works. Instead, we adopt two different centers, the dynamical H {\small I} center and the geometric center determined in \cite{ripepi17}. We find that the choice of center has an impact on the estimate of the COM motion, reflected in our systematic error. We find no compelling evidence for internal rotation, with a maximum rotation signal $V_{\mathrm{rot}} = 12 \pm 4$ km s$^{-1}$ when the H {\small I} center is used.
The resulting COM PMs roughly agree with all previously published values, though our random errors are several times smaller than the previous most precise measurement. This is primarily due to the increase in the size of the data set fit to the model. Soon after submission of this work, \cite{neiderhofer18} presented PMs from the VMC for a $3 \times 3$ degree region of the SMC. We do not attempt a detailed comparison here but their COM PM is marginally consistent with ours given the errors, however, they do not detect an outward residual motion toward the Bridge.

The small per-field errors allow us to probe the internal motions of the SMC, a galaxy whose internal structure is still quite unconstrained. We decompose the residual motion of each field (after subtraction of the COM motion) into a radial and a tangential component. We search for signs of rotation that would manifest as a signal in the tangential component as a function of distance from the SMC center. We see no clear trend in the tangential component.
We instead find evidence for large residual motions toward the east and west of the galaxy. The eastern residual motions, on the order of $\sim$80 km s$^{-1}$, point in the direction of the Magellanic Bridge. We estimate the escape speed from the SMC (see Section \ref{sec:fullstar}) and examine the impact of limiting the fields used in the COM PM calculation. We find that the removal of potentially unbound fields has little impact on the COM PM values. The areas of large residual motions also help explain the small differences in previous COM PM measurements, as both NK13 and \cite{cioni16} largely sampled the central and western regions of the SMC, which would not contain the significant residual motions seen in the eastern fields. 

This underscores the necessity of sampling a broad area of the SMC in determining a COM motion while also raising new questions about how to best build a model to fit the SMC moving forward. Previous LOS attempts to study SMC structure (e.g., \citealt{evans08}; \citealt{dobbie14a}) focused on the inner few degrees, where they did find a potential rotation signal. Our data set does not significantly probe the interior of the SMC, so we are unable to provide any further comparisons with these works. 
Finally, we also test whether different stellar populations in the SMC have measurable differences in their PMs. We employ the simple CMD cut shown in Figure \ref{fig:smccmd} to select a `blue' and a `red' stellar population, and re-derive the PMs for each field using these subsamples of stars. We do not find any statistically significant differences between the measured PMs for these two populations.

\subsection{Conclusions for Implied Orbit} \label{subsec:orbitconc}

Using the measured PMs, we find new Galactocentric velocities for the SMC and examine the consequences for its recent interaction history with the LMC. Our improved coverage of the SMC significantly improves our overall accuracy of the relative velocity between the two Clouds. 

Using this new relative velocity and two different mass cases for both the LMC and MW, we find a strong case for close interaction between the Clouds in the recent past (their centers of mass come within $\sim$20 kpc for 97\% of all cases examined). The mean COM distance is consistent within the errors across the two mass combinations that we consider, one with a heavy LMC and light MW (7.5$\pm$2.5 kpc 147$\pm$33 Myr ago), and a light LMC and a heavy MW (9.7$\pm$4.5 kpc 163$\pm$36 Myr ago), strongly supporting the idea of a direct collision between the Clouds. These impact parameters and encounter times depend little on our model for the internal PMs of the SMC
This lends support to the model of \citet{besla12}, where the Clouds have recently had a very close interaction, and where the LMC is thus primarily responsible (as opposed to the MW) for the morphology of the SMC, the Magellanic Stream and Bridge. This direct collision also has consequences for the morphology of the LMC.

\subsection{Future Work} \label{subsec:future}

We have presented an expanded picture of the PMs in the SMC, revealing its complicated dynamical nature. An immediate consequence of this is the necessity for a higher degree of spatial resolution. Improved resolution would help to disentangle where the ordered motion radially away from the
SMC begins and where there may be more coherent stellar rotation, if it exists in the SMC. Studies have shown an increasingly elongated picture of the SMC (e.g., \citealt{ripepi17}), so the combination of a higher PM spatial resolution with LOS studies could help create a data set that would have the power required to clearly identify the dynamical center of the SMC. The upcoming \textit{Gaia} Data Release 2 will provide the next opportunity.

We have better constrained the interaction history of the LMC and SMC. In future work, we will use this assumption to estimate the mass of the LMC enclosed within the SMC orbit. The mass of the LMC has been of considerable interest, first because it further constrains whether the Clouds are on their first or second passage about the MW, but it is also needed to ascertain the LMC's effect on the dynamics of the MW and its satellite population (e.g., \citealt{gomez15}; 
\citealt{penarrubia16}),  
and to better constrain how much debris came in with the LMC itself (e.g., \citealt{sales17}).

The direct collision between the Clouds that we discover here should also inform studies of star formation in the Clouds. We are able to determine a rough timescale for this encounter, and therefore correlations can be made between the past orbits of the Clouds and their star and cluster formation history. Already there is evidence that the locations and age gradients in the SMC star cluster population \citep{dias16} coincide well with the locations of our measured radial motions outwards in the outer regions, prima facie evidence that the ongoing interaction between the Clouds is inducing cluster formation. Future work combining these two datasets, the cluster population and SMC internal dynamics, provides a new opportunity to study the nature of star formation in an environment different than the posterchild Antennae Galaxies. 



\acknowledgments
Support for this work was provided by NASA through grants associated with projects GO-13476 and GO-14343 from the STScI, which is operated by the Association of Universities for Research in Astronomy, Inc., under NASA contract NAS 5-26555.
This work has made use of data from the European Space Agency (ESA) mission {\it Gaia} (\url{https://www.cosmos.esa.int/gaia}), processed by the {\it Gaia} Data Processing and Analysis Consortium (DPAC, \url{https://www.cosmos.esa.int/web/gaia/dpac/consortium}). Funding for the DPAC has been provided by national institutions, in particular the institutions participating in the {\it Gaia} Multilateral Agreement.
S.K. acknowledges the financial support of the Polish National Science
Center through the grants OPUS 2014/15/B/ST9/00093 and MAESTRO
2014/14/A/ST9/00121.

\bibliography{Zivick_bib}{}

\end{document}